\documentclass[12pt,preprint]{aastex}

\def\kms                 {km\thinspace s$^{-1}$}

\def\apjs{{\rm ApJS}}                  
\def\apjl{{\rm ApJ Let}}               

\def\Msol{\thinspace\hbox{$\hbox{M}_{\odot}$}}

\def\a4{\hsize 17.0cm \vsize 25.cm}
\newcommand{\der}[2]  { \frac{{\rm d}#1}{{\rm d}#2} }
\newcommand{\derp}[2] { \frac{\partial #1}{\partial #2} }
\newcommand{\dif}     {{\rm d}}
\newcommand{\difp}    {\partial}

\shorttitle{Star cluster winds}
\shortauthors{Silich et al.}

\begin{document}

\title{The Steady State Wind Model for Young Stellar Clusters with an 
           Exponential Stellar Density Distribution}

\author{Sergiy Silich \altaffilmark{1},
Gennadiy Bisnovatyi-Kogan \altaffilmark{2},
Guillermo  Tenorio-Tagle\altaffilmark{1}}
\and
\author{Sergio Mart\'{\i}nez-Gonz\'alez \altaffilmark{1}}
\altaffiltext{1}{Instituto Nacional de Astrof\'\i sica Optica y
Electr\'onica, AP 51, 72000 Puebla, M\'exico; silich@inaoep.mx}
\altaffiltext{2}{Space Research Institute, 84/32 Profsoyuznaya, Moscow,
                 117810, Russia; gkogan@iki.rssi.ru}

\begin{abstract}
A hydrodynamic model for steady state, spherically-symmetric winds driven 
by young stellar clusters with an exponential stellar density distribution is 
presented. Unlike in most previous calculations, the position of the 
singular point $R_{sp}$, which separates the inner subsonic zone from the 
outer supersonic flow, is not associated with the star cluster edge, but
calculated self-consistently. When the
radiative losses of energy are negligible, the transition from the subsonic
to the supersonic flow  occurs always at $R_{sp} \approx 4 R_c$, where 
$R_c$ is the characteristic scale for the stellar density distribution,
irrespective of other star cluster parameters. This is not the case in the
catastrophic cooling regime, when the temperature drops abruptly at a short 
distance from the star cluster center and the transition from the subsonic
to the supersonic regime occurs at a much smaller distance from the star 
cluster center.
The impact from the major star cluster parameters to the wind inner structure
is thoroughly discussed. Particular attention is paid to the effects which
radiative cooling provides to the flow. The results of the calculations 
for a set of input parameters, which lead to different hydrodynamic regimes,
are presented and compared to the results from non-radiative 1D numerical 
simulations and to those from calculations with a homogeneous stellar mass
distribution.
\end{abstract}


\section{Introduction}
\label{S1}

High spatial resolution observations have modified drastically our view of 
star formation
in starburst and normal galaxies leading to the discovery of a large number of
extremely massive and compact young stellar clusters with masses 
$10^5$\Msol - $10^6$\Msol, effective radii 3~pc - 5~pc and ages less 
than a few times 
$10^6$~yr - $10^7$~yr. The clusters, whose masses and expected lifetimes 
coincide with those of the globular clusters, and which are usually named 
in the literature super star clusters (see for a review \citealp 
{2009and..book..215T, 2010RSPTA.368..693D, 2010ARA&A..48..431P}).
Massive and compact star clusters are also a common feature in the nuclei of 
spiral galaxies. While the effective radii of nuclear clusters are similar to 
the sizes of globular and super star clusters, their luminosities and thus 
masses exceed them by up to two orders of magnitude 
(e.g. \citealp{2004AJ....127..105B, 2005ApJ...618..237W}).

Extremely high concentration (up to $10^4 - 10^5$~\Msol \, pc$^{-2}$) of young
stars implies that super star clusters (SSCs) are some of the most powerful 
feedback agents
which heat up, shape and enrich the interstellar gas with the product of 
supernova explosions. The combined action of such clusters may lead to the 
formation of powerful gaseous outflows (galactic superwinds) which link 
regions with extreme star  formation activity (starbursts) to the 
intergalactic medium (e.g. \citealp{1990ApJS...74..833H, 1995ApJ...438..563M,
2005ApJS..160...87R, 2008MNRAS.383..864W},
see for a review \citeauthor{2005ARA&A..43..769V} 2005
and references therein).
Thus, the hydrodynamic feedback from SSCs to the surrounding ISM is crucial in
many respects, as it is the origin and the duration of the starburst event,
the impact that starbursts provide to their host galaxies and the
intergalactic medium, the evolution of the assembling galaxies
and the feeding of supermassive black holes located within nuclear
starburst regions.

It was suggested by \citet{1985Natur.317...44C} (CC85)
and \citet{1992ApJ...397L..39C}
that the kinetic energy supplied by supernovae and stellar winds is 
thermalized within the star cluster volume. This leads to a high central 
overpressure which drives the inserted gas away from the cluster in the form 
of a star cluster wind. If the sources of energy and mass (stars) are 
homogeneously distributed inside the star cluster volume with an outer radius 
$R_{SC}$ and the radiative losses of energy are negligible, then the 
hydrodynamics of the outflow depend only on the energy and mass deposition 
rates, $L_{SC}$ and ${\dot M}_{SC}$, and the radius of the cluster $R_{SC}$. 
In this case the distributions of the hydrodynamical variables can be 
calculated analytically 
(see \citealp{1985Natur.317...44C, 2000ApJ...536..896C}).
\citet{2003ApJ...590..791S, 2004ApJ...610..226S},
\citet{2007ApJ...658.1196T}, \citet{2008ApJ...683..683W, 2011ApJ...740.75W} 
showed that in the case of very massive and compact clusters, 
radiative cooling becomes an important issue and developed a radiative star 
cluster wind model.

Real clusters, however, are not homogeneous. Observations (see  
\citealp{2010ARA&A..48..431P}
and references therein) show that the
stellar density drops rapidly with distance to the star cluster center
and that the characteristic scale for the stellar density distribution
(core radius) is much smaller then the star cluster effective radii.
This effect has been discussed by \citet{2007MNRAS.380.1198R}
who found an analytic non-radiative solution in the case of power law
stellar density distributions and then compared it to 3D gas dynamic 
simulations. It is important to note that in 
\citet{2007MNRAS.380.1198R} solution the central gas density goes to infinity 
if the stellar density drops faster than $1/r$. This implies that for steep
stellar density distributions radiative cooling in the central zone of the 
cluster might also be a crucial factor. Note also that in
\citet{2007MNRAS.380.1198R} solution the stellar density 
(and thus the energy and the mass deposition rates) abruptly drop to zero 
value at some distance from the star cluster center and it is assumed that
the transition from the subsonic to supersonic regime occurs exactly at
this point. Thus, in 
\citet{2007MNRAS.380.1198R} model the 
cut-off radius in the stellar density distribution remains an important 
parameter although the stellar density distribution is not homogeneous. 
\citet{2006MNRAS.372..497J} neglected the gravitational pull from the 
cluster and 
radiative cooling and integrated the 1D stationary hydrodynamic equations
numerically in the case of an exponential stellar density distribution. 
These results were then used to discuss the impact of  
non-equilibrium ionization onto the star cluster wind X-ray emission. 

Here we present a steady state semi-analytic hydrodynamic solution for winds 
driven by stellar clusters with an exponential stellar density distribution, 
which includes also radiative cooling. As in the above mentioned papers, 
we regard such a stellar distribution as more realistic than the formerly
used homogeneous one. We  thoroughly discuss boundary
and initial conditions, which define the position of the singular point, and
thus the radius where the flow changes the hydrodynamic regime and becomes
supersonic. We also discuss the impact which radiative cooling
provides to the flow. The paper is organized as follows: 
the input star cluster wind model is formulated in section 2. In section 3 
we introduce the set of main equations and present them in the form
suitable for numerical integration. The initial and boundary conditions
are discussed in section 4. The method, which allows to select a wind 
solution from a branch of possible integral curves, is described
in section 5. In section 6 we first present the results of the simulations
for three reference models with different star cluster mechanical 
luminosities and then discuss how other input parameters affect the
solution. An extreme regime with catastrophic gas cooling is discussed 
in section 7. We compare our results with those obtained under the
assumption that stars are homogeneously distributed within the cluster volume 
in section 8. Our major results are summarized in section 9. 

\section{The model}
 \label{S2}

We consider a young stellar cluster with constant energy and mass 
deposition rates $L_{SC}$ and ${\dot M}_{SC}$ and an exponential stellar 
density distribution:
\begin{equation}
      \label{eq1}
\rho_{\star}(r) = \rho_{\star 0} \exp(-r/R_c) ,
\end{equation}
where $\rho_{\star 0}$ is the central stellar density and
$R_c$ is the radius of the star cluster ``core''. The total mass of the 
cluster is then:
\begin{equation}
      \label{eq2}
M_{SC} = 4 \pi \int_0^{\infty} \rho_{\star 0} r^2 \exp(-r/R_c) \dif r = 
         8 \pi \rho_{\star 0} R^3_c . 
\end{equation}
Note, that the half-mass radius of the cluster in the case of exponential 
stellar density distribution is: $R_{hme} = 2.67 R_c$. 

It is assumed that the kinetic energy supplied by stellar winds and supernova 
explosions is completely thermalized, that the gravitational field of the 
cluster is negligible, and that the energy and mass deposition rates per 
unit volume, $q_e$ and $q_m$, are in direct proportion to the local stellar 
density: 
\begin{eqnarray}
      \label{eq.2a}
      & & \hspace{-1.1cm} 
q_e(r) = q_{e0} \exp(-r/R_c) ,
      \\[0.2cm]     \label{eq.2b}
      & & \hspace{-1.1cm}
q_m(r) = q_{m0} \exp(-r/R_c) ,
\end{eqnarray}
where the normalization constants $q_{e0}$ and $q_{m0}$ are:
\begin{eqnarray}
      \label{eq.3a}
      & & \hspace{-1.1cm} 
q_{e0} = L_{SC} / 8 \pi R^3_c ,
      \\[0.2cm]     \label{eq.3b}
      & & \hspace{-1.1cm}
q_{m0} =  {\dot M}_{SC} / 8 \pi R^3_c .
\end{eqnarray}

\section{Basic equations}
\label{S3}

The hydrodynamic equations for the stationary, spherically symmetric
flow driven by stellar winds and supernova explosions if
gravity is neglected are (see, for example, \citealp{1971ApJ...165..381J,
1985Natur.317...44C, 2000ApJ...536..896C, 2004ApJ...610..226S, 
2006MNRAS.372..497J}):
\begin{eqnarray}
      \label{eq4a}
      & & \hspace{-1.0cm}
\frac{1}{r^2} \der{}{r}\left(\rho u r^2\right) = q_m ,
      \\[0.2cm]
      \label{eq4b}
      & & \hspace{-1.0cm}
\rho u \der{u}{r} = - \der{P}{r} - q_m u ,
      \\[0.2cm]
     \label{eq4c}
      & & \hspace{-1.0cm}
\frac{1}{r^2} \der{}{r}{\left[\rho u r^2 \left(\frac{u^2}{2} +
\frac{\gamma}{\gamma - 1} \frac{P}{\rho}\right)\right]} = q_e - Q ,
\end{eqnarray}
where $u$, $P$, and $\rho$ are the outflow velocity, thermal pressure
and density, $\gamma$ is the ratio of the specific heats,
$Q = n_e n_i \Lambda(T,Z)$ is the cooling rate and 
$\Lambda(T,Z)$ is the cooling function, which depends on the gas
temperature $T$ and metallicity $Z$. Hereafter we relate the
energy and the mass deposition rates, $L_{SC}$ and ${\dot M}_{SC}$,
via the equation:
\begin{equation}
      \label{eq5}
L_{SC} = {\dot M}_{SC} V^2_{A\infty} / 2 ,
\end{equation}
and assume that the adiabatic wind terminal speed, $V_{A\infty}$, is 
constant. Thus, the model input parameter $V_{A\infty}$ defines the
mass deposition rate for a given cluster mechanical luminosity
$L_{SC}$.

The integration of the mass conservation equation (\ref{eq4a}) yields:
\begin{equation}
      \label{eq6}
\rho u r^2 = - q_{m0} (R_c r^2 + 2 R_c^2 r + 2 R_c^3) \exp(-r/R_c) + C \, .
\end{equation}
If the density and the velocity of the flow in the star cluster center
are finite, the constant of integration is: $C = 2 q_{m0} R_c^3$.  
Using this expression and taking the derivative of equation 
(\ref{eq4c}), one can present the main equations in the form:
\begin{eqnarray}
 \label{eq7a}
      & & \hspace{-1.1cm} 
\der{u}{r}  = \frac{(\gamma-1)(q_e - Q) + (\gamma+1) q_m u^2 / 2 -
              2 \rho u c^2/r}{\rho (c^2 - u^2)} 
      \\[0.2cm]     \label{eq7b}
      & & \hspace{-1.1cm}
\der{P}{r} = - \rho u \der{u}{r} - q_m u \, ,
      \\[0.2cm] \label{eq7c}
      & & \hspace{-1.1cm}
\rho = \frac{2 q_{m0} R_c^3}{r^2 u} \left[1 - \left(1 +  \frac{r}{R_c} + 
       \frac{1}{2}\frac{r^2}{R^2_c}\right)
       \exp(-r/R_c)\right]  \, ,
\end{eqnarray}
where $c$ is the local speed of sound, $c^2 = \gamma P / \rho$.  
Note that the central density remains finite and is not zero, 
$\rho \ne 0$~g cm$^{-3}$, only if the 
flow velocity in the star cluster center is 0~\kms and grows linearly with 
radius near the center. The derivatives of the wind 
velocity and pressure in the star cluster center then are:
\begin{eqnarray}
 \label{eq8a}
      & & \hspace{-1.1cm} 
\der{u}{r} = \left[(\gamma-1)(q_e - Q) - 2 q_{m0} c^2 / 3 \right]  / 
      \rho  c^2 \,  ,
      \\[0.2cm]     \label{eq8b}
      & & \hspace{-1.1cm}
\der{P}{r} = 0 \, ,
\end{eqnarray}
We make use of these equations in order to move away from the center and
start the numerical integration.

\section{Initial and boundary conditions}
\label{S4}

In the non-radiative solution the sound speed in the center is defined 
directly by the adiabatic wind terminal speed $V_{A\infty}$ and does not 
depend on the wind central density  (CC85,  Cant{\'o} et al. 2000): 
$c_{A0} = [(\gamma - 1) / 2]^{1/2} V_{A\infty}$. This is  not the
case when radiative cooling is taken into consideration. The central gas 
density, $\rho_c$, and the central temperature, $T_c$, are then related by 
the equation (\citealp{1987ApJ...320...32S, 2004ApJ...610..226S}):
\begin{equation}
      \label{eq9}
\rho_{c} = q_{m0}^{1/2}\left[\frac{V_{A,\infty}^{2}/2 - 
            c_{0}^{2}/(\gamma -1)}{\Lambda (Z,T_{c})}\right]^{1/2} \, ,
\end{equation}
where $c_0$ is the sound speed in the star cluster center.
Equation (\ref{eq9}) shows that the central temperature in the radiative 
solution can never exceed the non-radiative value, 
$T_{c0} = \mu_a c^2_{c0} / \gamma k$, where $k$ is the Boltzmann's constant 
and $\mu_a = 14/23 m_H$ is the mean mass per particle in the fully ionized 
plasma with 1~He atom per 10~H atoms. It also defines the thermal 
pressure in the star cluster center if the central temperature is known:
\begin{equation}
      \label{eq10}
P_{c} = k n_{c} T_{c} = k \rho_{c} T_{c} / \mu_a \, .
\end{equation}
Thus, the value of the central temperature selects the unique 
solution from the branch of possible integral curves presented in 
Figure 1. 
\begin{figure}[htbp]
\plotone{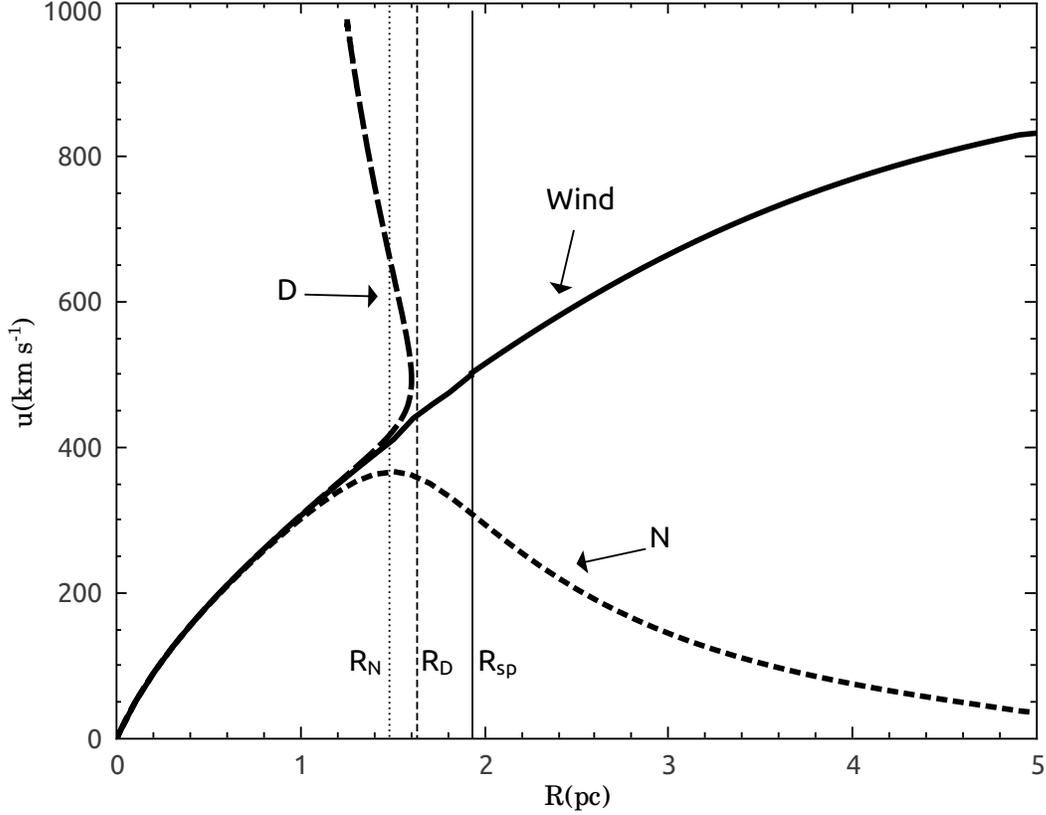}
\caption{Possible integral curves. The value of the central temperature
selects the type of the integral curve. If the central temperature is
large, the integral curve reaches the maximum radius $r = R_D$ (vertical 
dashed line), where the denominator in equation (\ref{eq7a}) changes sign 
and the velocity becomes supersonic. The integral curve then goes back towards
the center. This is non physical double value solution. If the central 
temperature is low, the integral curve reaches the maximum velocity at 
$r = R_N$ (vertical dotted line), where the numerator in equation 
(\ref{eq7a}) changes the sign. The velocity then drops with distance to the 
star cluster center. The velocity is always subsonic in this case. This is 
known as a ``breeze'' solution (e.g. \citealp{1995ApJ...453..369B}).
The wind solution is 
the unique solution, which passes by the singular point at $r = R_{sp}$ 
(vertical solid line), where both, the numerator and denominator in 
equation (\ref{eq7a}) vanish and the flow becomes supersonic for larger 
radius.}
\label{fig1}
\end{figure}
The wind solution is the unique solution, which passes through the singular
point, where both, the numerator and the denominator in equation (\ref{eq7a}),
vanish and the subsonic flow in the inner zone, $r < R_{sp}$, becomes 
supersonic at $r > R_{sp}$. This outer boundary condition i.e. that the 
integral curve must pass through the singular point, allows one to select the 
value of the central temperature which leads to the wind solution, and also 
defines the singular point position.

\section{The integration procedure}
\label{S5}

As shown in Figure 1, there are three possible types of integral curves. 
We will call them N-type, D-type and wind-type solutions, as 
indicated in Figure 1. The N-type solution is subsonic everywhere. In this
case the velocity grows monotonically with distance from the star
cluster center until it reaches a maximum value when the numerator in 
equation (\ref{eq7a}) vanishes. For larger distances the numerator in 
(\ref{eq7a}) becomes negative, however the denominator does not, and thus the 
velocity drops and the flow remains always subsonic. The D-type solution 
passes the sonic point at some distance from the center. At this point the 
denominator in (\ref{eq7a}) vanishes, however the numerator does not.
The integral curve then reaches the maximum radius and then turns back
towards the center as the velocity grows. Note that one can
easily obtain this solution, if the velocity is used instead of the radius as 
the independent variable. The wind-type solution is the only solution,
which meets the singular point where both, the numerator and 
denominator in the equation of motion vanish. 

Which type of the integral curve is selected, depends on the value of the 
central temperature.
D-type solution occurs when the central temperature exceeds that 
which results into the wind-type solution. N-type solution occurs in the
opposite case. This allows to build up a simple iteration procedure, which
allows one to obtain the central temperature and the position of the singular 
point for the wind-type solution. The procedure is based on the bisection
method and includes three integrations with three different central 
temperatures, $T_{max}$, $T_{min}$ and $T_A = (T_{max} + T_{min}) / 2$ at each 
iteration step. The initial values of $T_{max}$ and $T_{min}$ must be 
selected in
such a way that $T_{max}$ is larger and $T_{min}$ is smaller than the
wind-type central temperature. The central temperature, $T_c$, for the 
wind-type solution is always between the values of 
$T_{max}$ and $T_{min}$: $T_{min} < T_c < T_{max}$.
At the end of each iteration step, either $T_{max}$ or $T_{min}$ is replaced 
with $T_A$, what allows to narrow the interval for the
wind central temperature and approach closer to the singular point. We 
usually stop iterations, when the position 
of the singular point is within an accuracy $\delta \le 10^{-3}$, where
$\delta = [(R_{A}-R_{min})^2 + (R_{A} - R_{max})^2]^{1/2} / R_A$.
$R_{max}$, $R_{min}$ and $R_A$ are the radii, at which the denominator or 
numerator in the equation of motion (\ref{eq7a}) vanishes in the solutions 
with central temperatures $T_{max}$, $T_{min}$ and $T_A$, respectively. This 
procedure allows one to obtain the value of the wind central 
temperature and to localize the position of the singular point with 
high accuracy. 

However, it does not allow to pass through the singular point and obtain the 
runs of the hydrodynamic variables for $r \ge R_{sp}$. In order to extend 
integral curves outside of the singular radius $R_{sp}$ and complete
the solution, one must know the values and the derivatives of the 
hydrodynamical variables in the singular point. We obtain these quantities 
from the condition that at the singular point both, the numerator and the 
denominator in equation (\ref{eq7a}) vanish and thus the velocity of the flow 
is equal to the local speed of sound, $u_{sp} = c_{sp}$ (see Appendix). 

\section{Results and discussion}
\label{S6}

In order to verify our model, we first compared our results with those obtained
by Ji et al. (2006), who integrated numerically equations 
(\ref{eq4a})-(\ref{eq4c}) assuming that the impact from radiative cooling is 
negligible. Ji et al. (2006) obtained the singular radius $R_{sp} = 1.97$~pc 
for $R_c = 0.48$~pc, ${\dot M}_{SC} = 10^{-4}$~\Msol \, yr$^{-1}$ and
$V_{A\infty} = 1000$~km s$^{-1}$ (see section 4 and Figure 8 in their 
paper). The run of the wind velocity obtained with our code for such an input 
model is shown in Figure 1 by the solid line. It is very similar to that,
obtained in the 1D simulations (note that Ji et al. normalized their radii
to the singular radius, $R_{sp}$). The radius of the singular point in our
calculations is $R_{sp} = 1.94$~pc. This implies that in the quasi-adiabatic
case our model is in excellent (about 1.5\%) agreement with the 1D results of
\citet{2006MNRAS.372..497J}.
\begin{table}[htp]
\caption{\label{tab1} Reference models}
\begin{tabular}{c c c c c c}
\hline\hline 
Model  & Core radius & Half-mass radius & Mechanical luminosity & Adiabatic terminal speed \\
       & (pc)          & (pc)             & (erg s$^{-1}$)   & (km s$^{-1}$) \\
\scriptsize{(1)} & \scriptsize{(2)} & \scriptsize{(3)} &\scriptsize{(4)} 
                 & \scriptsize{(5)} &  \\
\hline
A  &  1.0 & 2.67  & $3 \times 10^{40}$ & 1000 &  \\
B  &  1.0 & 2.67  & $3 \times 10^{39}$ & 1000 &  \\ 
C  &  1.0 & 2.67  & $3 \times 10^{41}$ & 1000 &  \\ 
\hline\hline
\end{tabular}
\end{table}

Figure 2 shows the results of the calculations for our three reference
models presented in Table 1. The model clusters have the same core radius 
(1~pc) and adiabatic wind terminal speed (1000~km~s$^{-1}$) but different 
mechanical luminosities: $L_{SC} = 3 \times 10^{40}$~erg s$^{-1}$, 
$L_{SC} = 3 \times 10^{39}$~erg s$^{-1}$ and  
$L_{SC} = 3 \times 10^{41}$~erg s$^{-1}$. These mechanical luminosities 
correspond to young stellar clusters with masses about $M_{SC} = 10^6$\Msol, 
$10^5$\Msol \, and $10^7$\Msol, respectively. The distributions of 
velocity, temperature, density and 
pressure are shown in panels a, b, c and d, respectively. The solid, dotted 
and dashed lines correspond to cases A, B and C. In all cases the 
flow velocity near the center grows almost linearly with radius,
passes the singular point at about 4~pc distance from 
the center and then gradually approaches the terminal speed value.
The wind velocities are almost identical in the quasi-adiabatic models
A and B (solid and dotted lines, respectively). 

The impact of strong radiative cooling is noticeable in the most energetic 
case C. In this case the terminal wind velocity is smaller than in the 
cases A and B because radiative cooling reduces the amount of thermal energy 
in the central zone, that results into a smaller wind energy, and thus 
smaller terminal speed. The impact from radiative cooling is best noticed in 
the radial profiles of temperature and thermal pressure (panels b and 
d, respectively). 
In the quasi-adiabatic cases A and B the temperature drops slowly with  
distance from the star cluster core showing almost un-distinguishable 
distributions (solid and dotted lines in panel b). 
In model C the temperature profile is different. The temperature deviates  from
the quasi-adiabatic profile significantly already at about ten times the core 
radius and then drops rapidly to the lower permitted value, of about 
$10^4$~K, at about 35~pc from the center. This leads to the fast decrease in 
the thermal pressure (panel d) which drops more than an order of magnitude
at this distance. Note, that thermal pressure always drops rapidly with 
distance from the star cluster center in the free wind region. 
Figure 2  shows also 
that the central pressure grows with the star cluster mass/power. This is 
also the case for the central density (see panel c). 
However, radiative cooling does not affect the density distribution 
significantly even in the most powerful case C (panel c). 
\begin{figure}[htbp]
\vspace{18.0cm}
\includegraphics{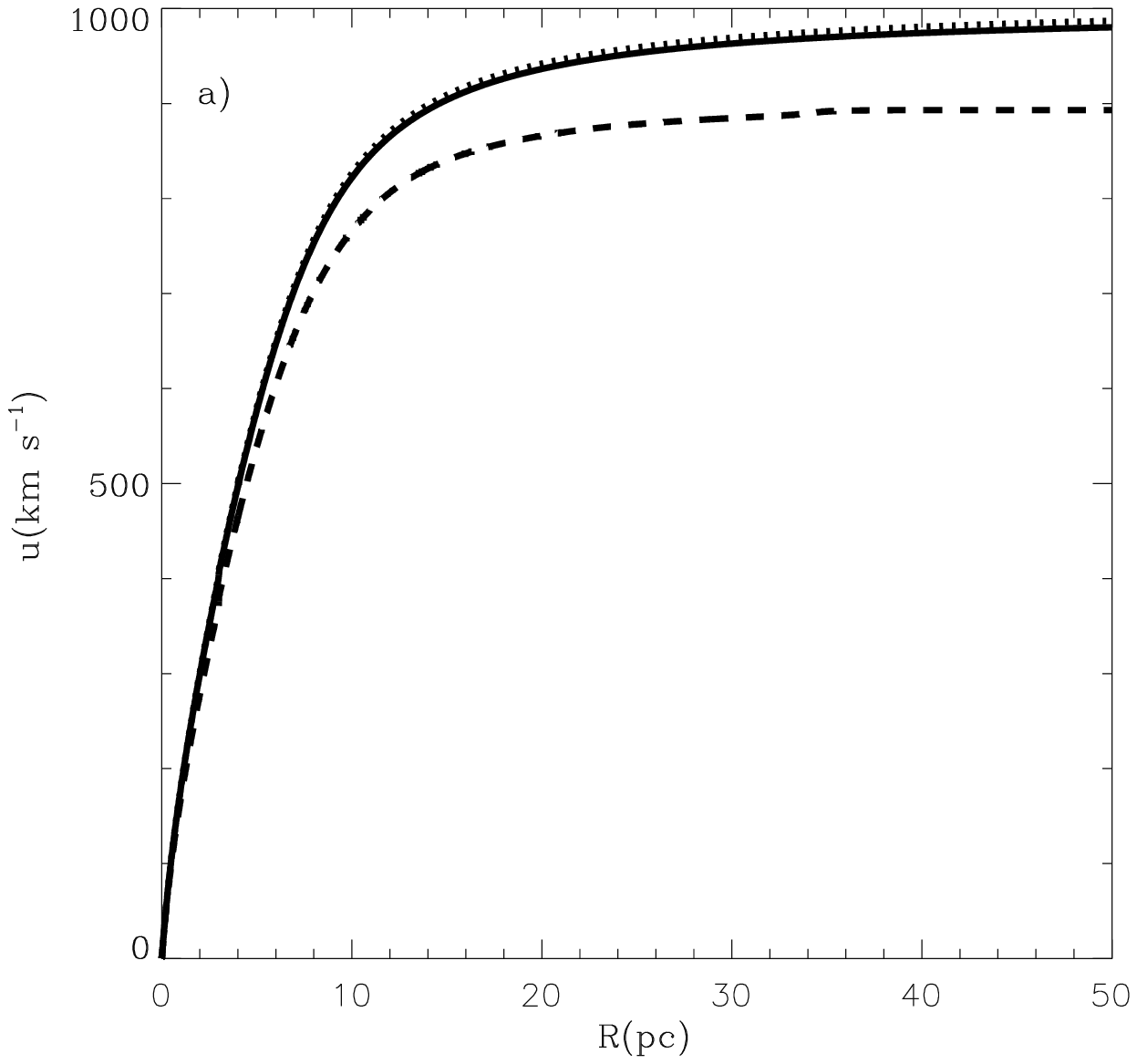}
\includegraphics{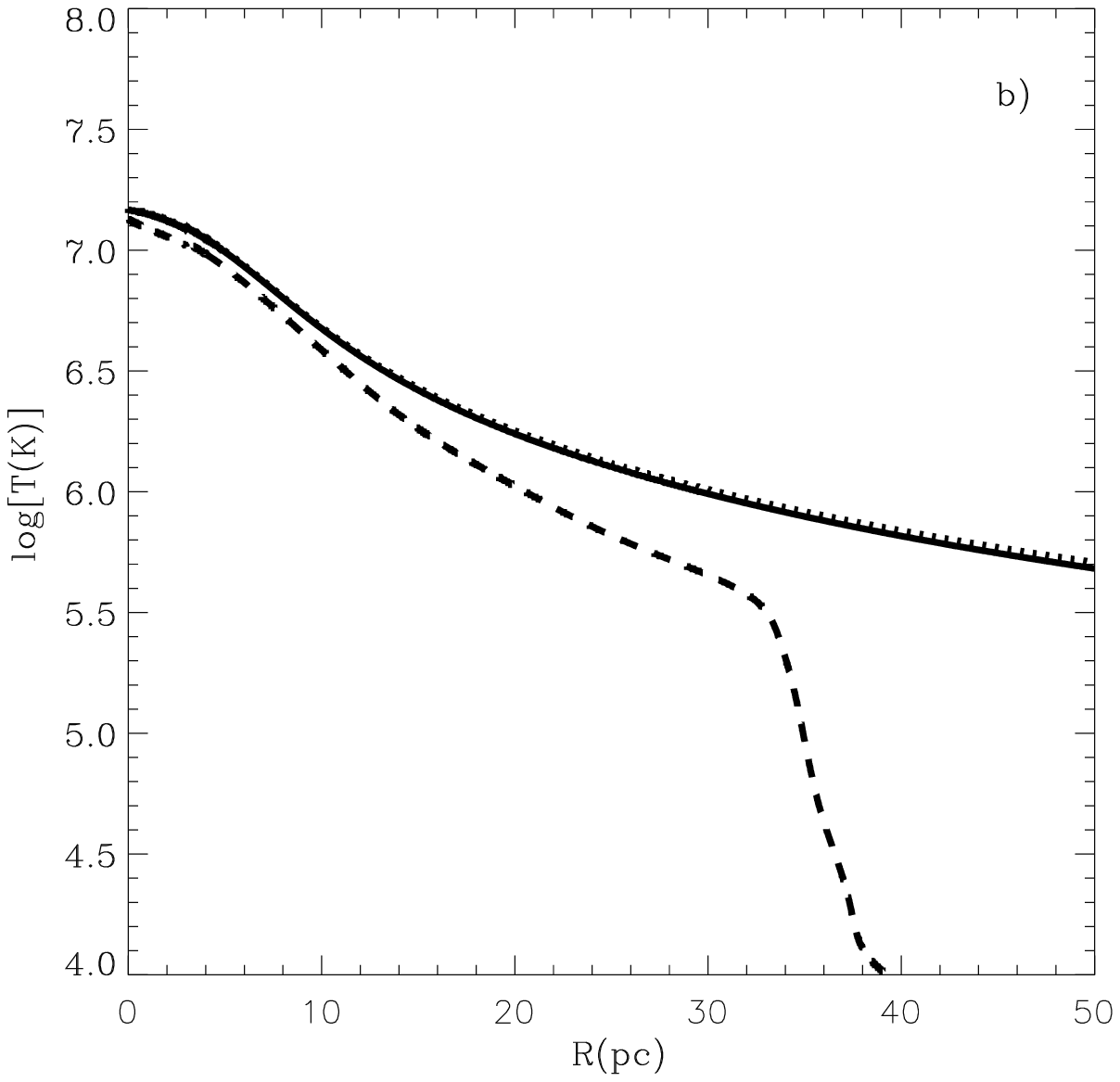}
\includegraphics{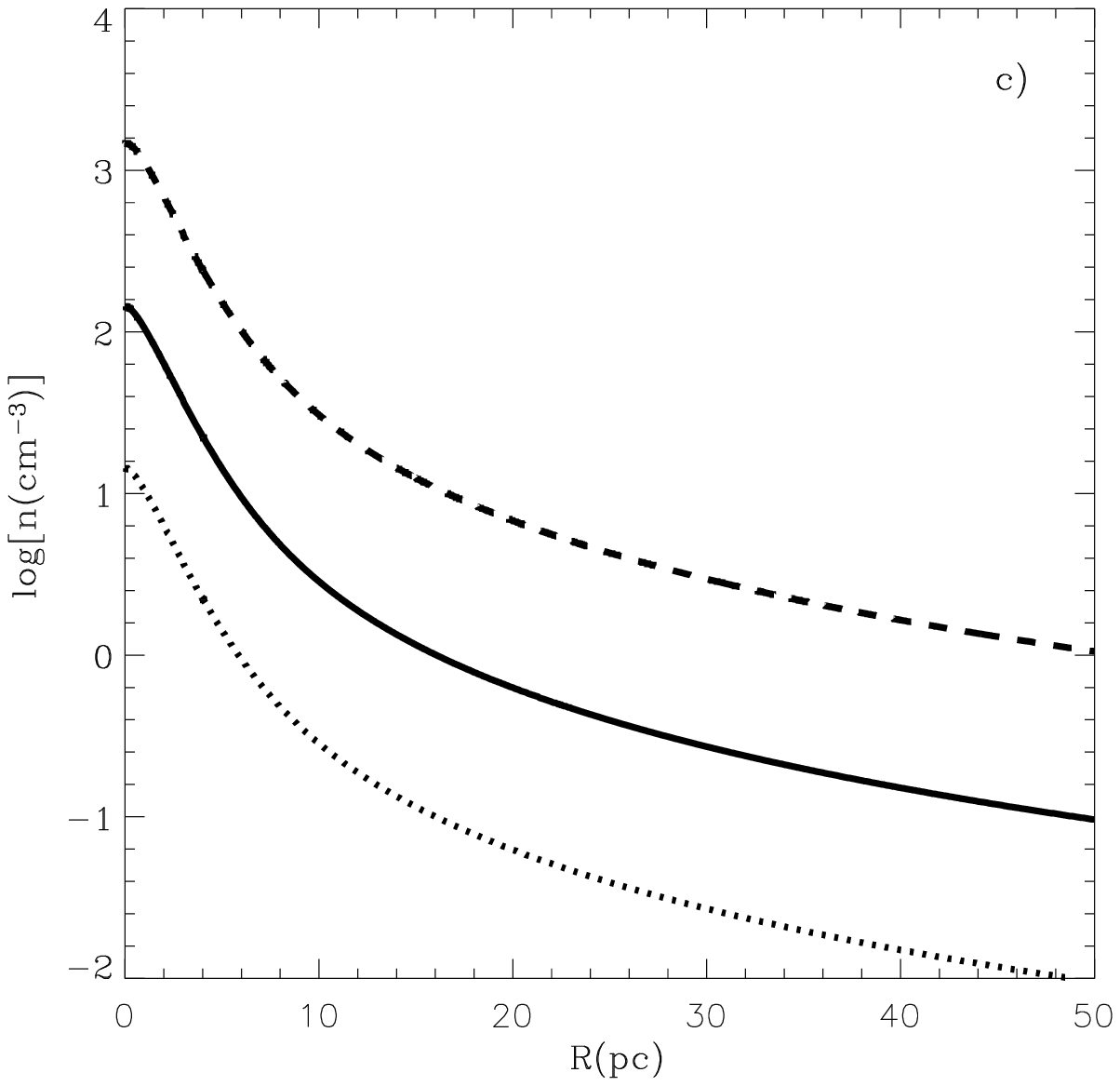}
\includegraphics{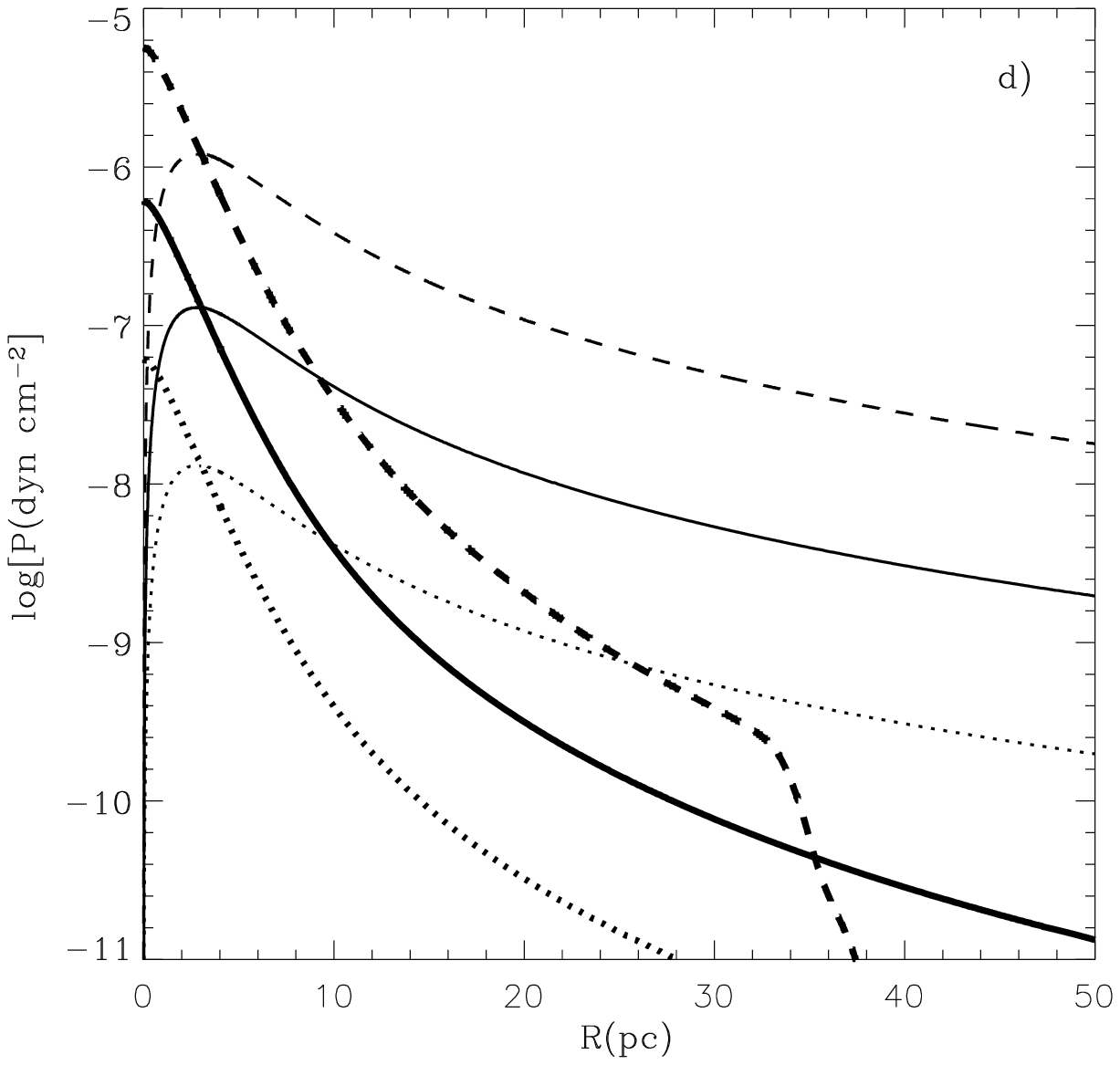}
\caption{The stationary wind solution. Panels a, b, c and d present the
wind velocity, temperature, density and pressure, respectively. 
Solid, dotted and dashed lines present the results of the calculations
for models A, B and C. Thick and thin lines in panel d display the 
thermal and ram pressure, respectively.}
\label{fig2}
\end{figure}

The value of $V_{A\infty}$ parameterizes the ratio of the star cluster 
mechanical luminosity to the mass deposition rate. Both parameters,
$L_{SC}$ and ${\dot M}_{SC}$, vary as the cluster evolves (e.g. 
\citealp{1999ApJS..123....3L}).
Beside this, the flow may be mass loaded by the gas left over 
from star formation 
(\citealp{2003MNRAS.339..280S, 2010ApJ...711...25S}).
Figure 3 presents the results of the calculations for three different
values of this parameter: $V_{A\infty} = 1000$~km s$^{-1}$, $V_{A\infty} = 
750$~km s$^{-1}$ and $V_{A\infty} = 2000$~km s$^{-1}$ (solid, dashed and
dotted lines, respectively). The rest of the input parameters in this set 
of models are the same as in the reference model A: 
$L_{SC} = 3 \times 10^{40}$~erg s$^{-1}$ and $R_c = 1$~pc.
The calculated wind central temperature (see panel b) increases for
larger $V_{A\infty}$ as it is also the case in the non-radiative solution 
(see  Cant{\'o} et al. 2000). The calculated wind terminal speed is 
similar to the adiabatic value when $V_{A\infty}$ parameter is large.
However, in the low velocity case, $V_{A\infty} = 750$~km s$^{-1}$, the 
difference between the adiabatic and the calculated terminal speeds is 
noticeable. This is because the density in the wind, $\rho = {\dot M}(r) / 
4 \pi r^2 u(r)$, is larger if the selected adiabatic wind terminal speed 
is smaller (Figure 3, panel c) that leads to a stronger radiative cooling. 
Strong radiative cooling removes a fraction of the deposited energy and  
thus decreases the flux of mechanical energy and consequently the wind 
terminal speed. Indeed,
in the quasi-adiabatic cases with  $V_{A\infty} = 2000$~km s$^{-1}$ and
$V_{A\infty} = 1000$~km s$^{-1}$ the amount of radiated energy is negligible,
but it reaches about $14\%$ of the deposited energy in the model
with  $V_{A\infty} = 750$~km s$^{-1}$.
In this case, radiative cooling leads to the rapid decrease in the temperature 
distribution at about 35~pc distance from the star cluster center as it is 
also the case in the reference model C. This results into the sharp drop of 
thermal pressure at the same distance (see panel d). Nevertheless, the 
position of the singular point, $R_{sp}$, remains about the same, 
$R_{sp} \approx 4$~pc.
\begin{figure}[htbp]
\vspace{18.0cm}
\includegraphics{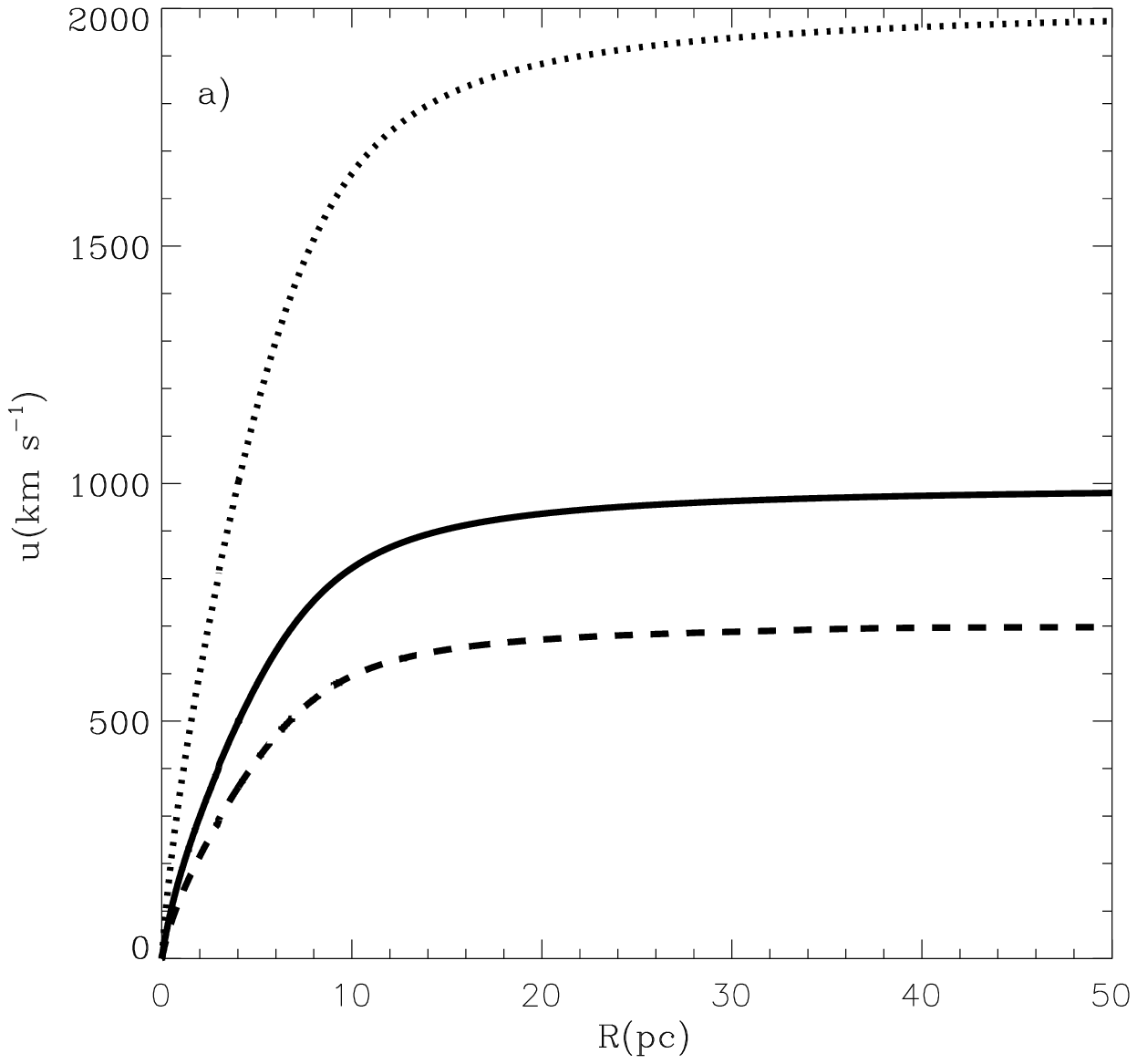}
\includegraphics{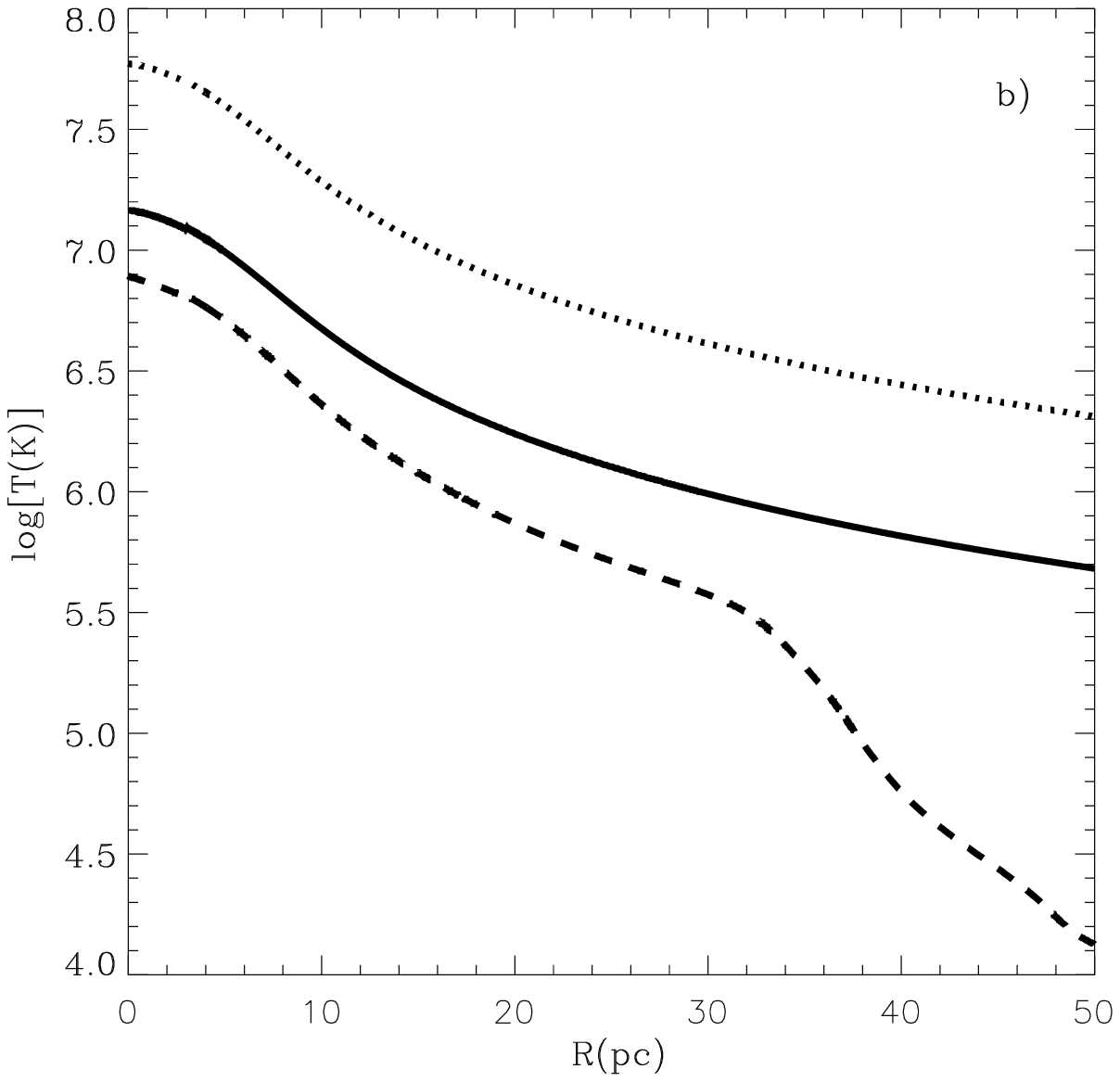}
\includegraphics{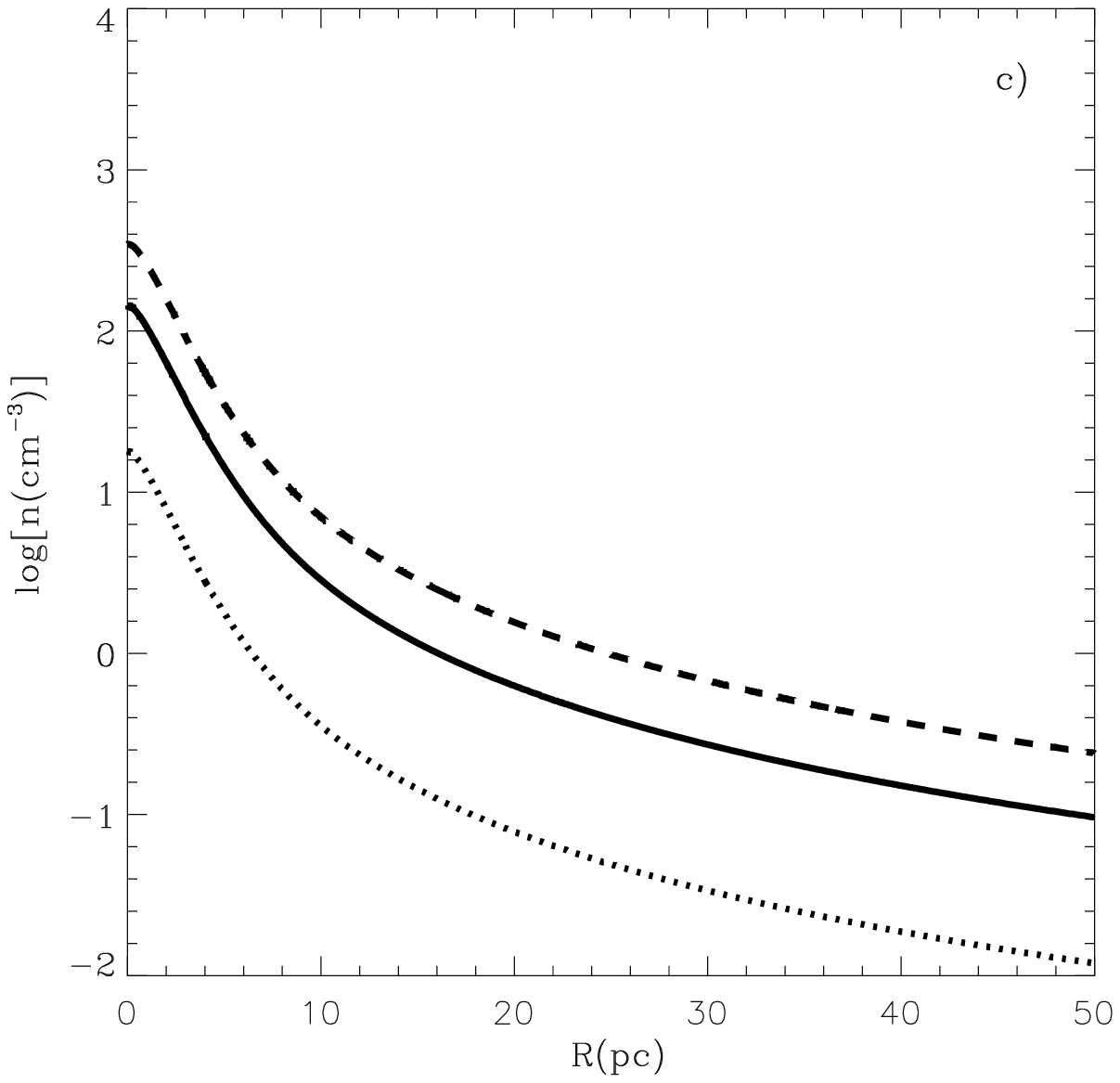}
\includegraphics{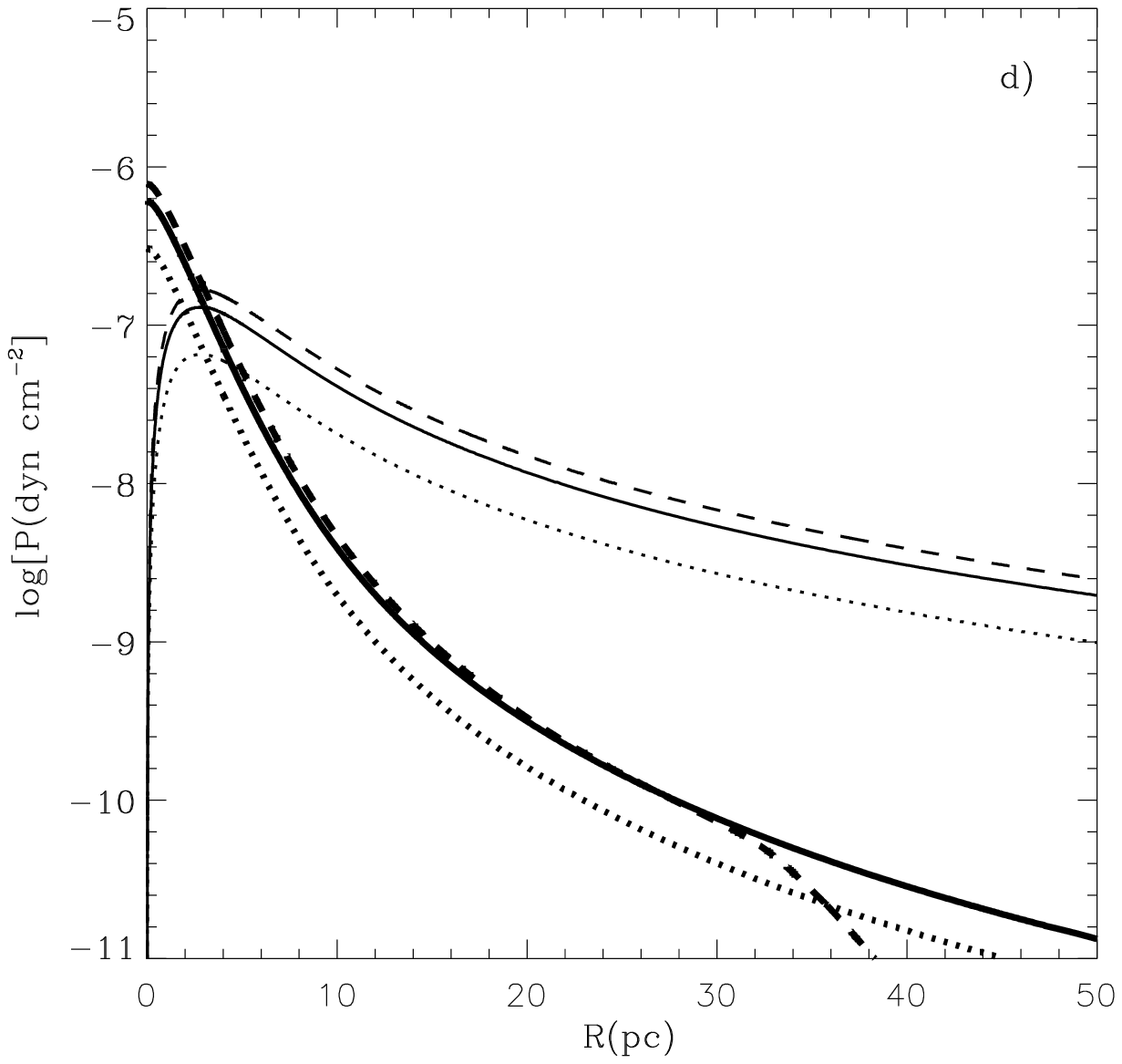}
\caption{The results of the calculations for models with
different $V_{A\infty}$. Panels a, b, c and d present the
wind velocity, temperature, density and pressure, respectively. 
Solid, dotted and dashed lines present the results of the calculations
for models with $V_{A\infty} = 1000$~km s$^{-1}$, $V_{A\infty} = 
2000$~km s$^{-1}$ and $V_{A\infty} = 750$~km s$^{-1}$, respectively.
Thick and thin lines in panel d display the thermal and ram 
pressure, respectively.}
\label{fig3}
\end{figure}

Finally, Figure 4 shows how the distributions of the hydrodynamical variables
change with the star cluster core radius. This figure displays the resulting
velocity, temperature, density and pressure profiles for clusters which have 
the same
mechanical luminosity ($L_{SC} = 3 \times 10^{40}$~erg s$^{-1}$) and adiabatic
wind terminal speed ($V_{A\infty} = 1000$~km s$^{-1}$), but different 
core radii: $R_c = 1$~pc,  $R_c = 5$~pc and  $R_c = 0.2$~pc (solid, dotted and 
dashed lines, respectively). Certainly, the speed of the wind grows faster
when the cluster is more compact, as it is shown in panel a.
The central temperatures are the same in all cases.
However, the temperature profiles are different. In the case with a larger 
core radius the flow is quasi-adiabatic and the temperature drops slowly with 
radius. On the other hand, the most compact model with
$R_c = 0.2$~pc is strongly affected by radiative cooling (dashed
line in panel b). In this case the temperature drops rapidly and reaches
the $10^4$~K value at about 17~pc distance from the center.
One can also notice the effects of strong radiative cooling on panel d, which 
displays the calculated star cluster wind thermal and ram 
pressure profiles. The value of the core radius affects also the wind central 
density which grows for smaller core radius, as shown in panel c.
\begin{figure}[htbp]
\vspace{18.0cm}
\includegraphics{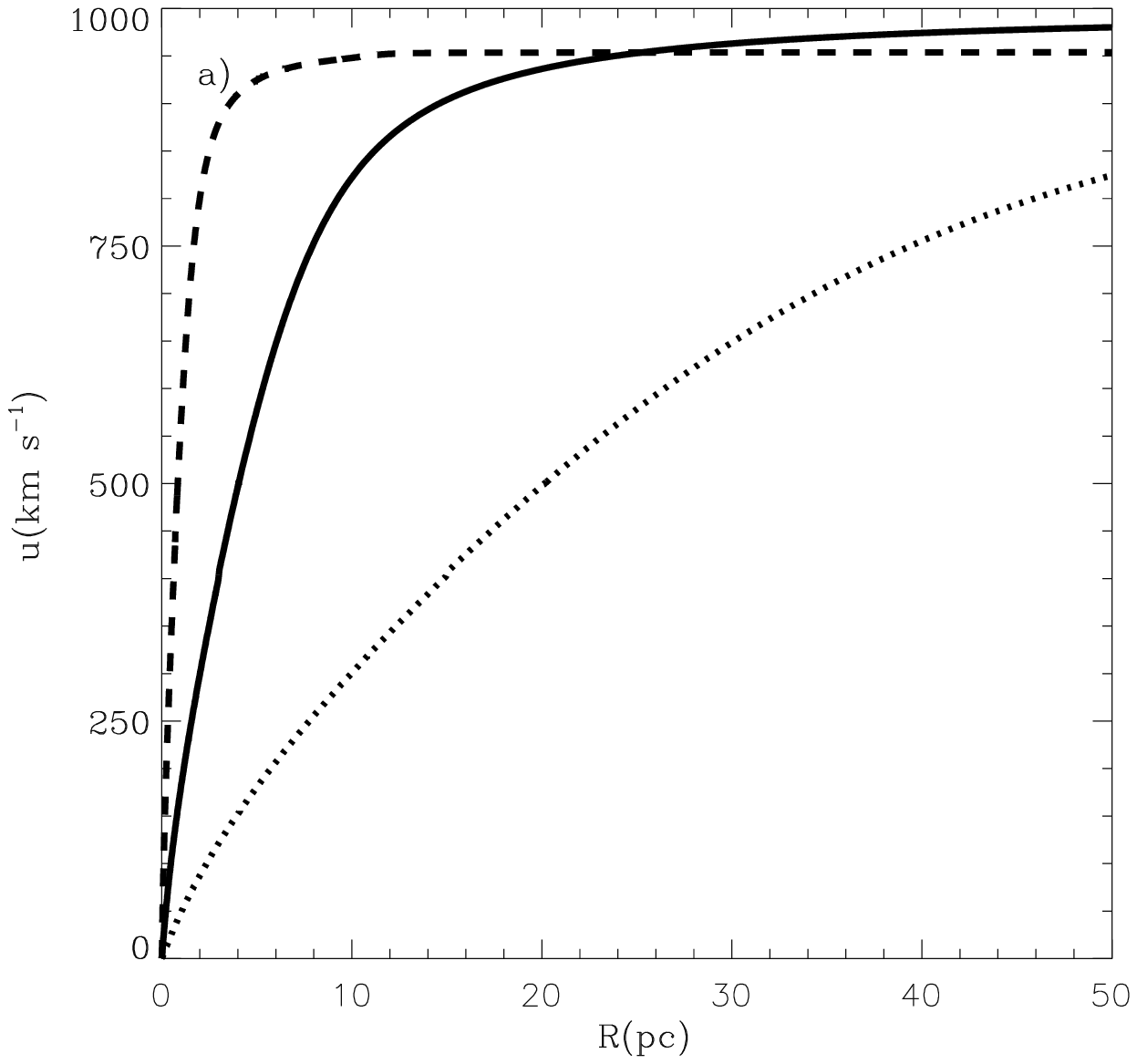}
\includegraphics{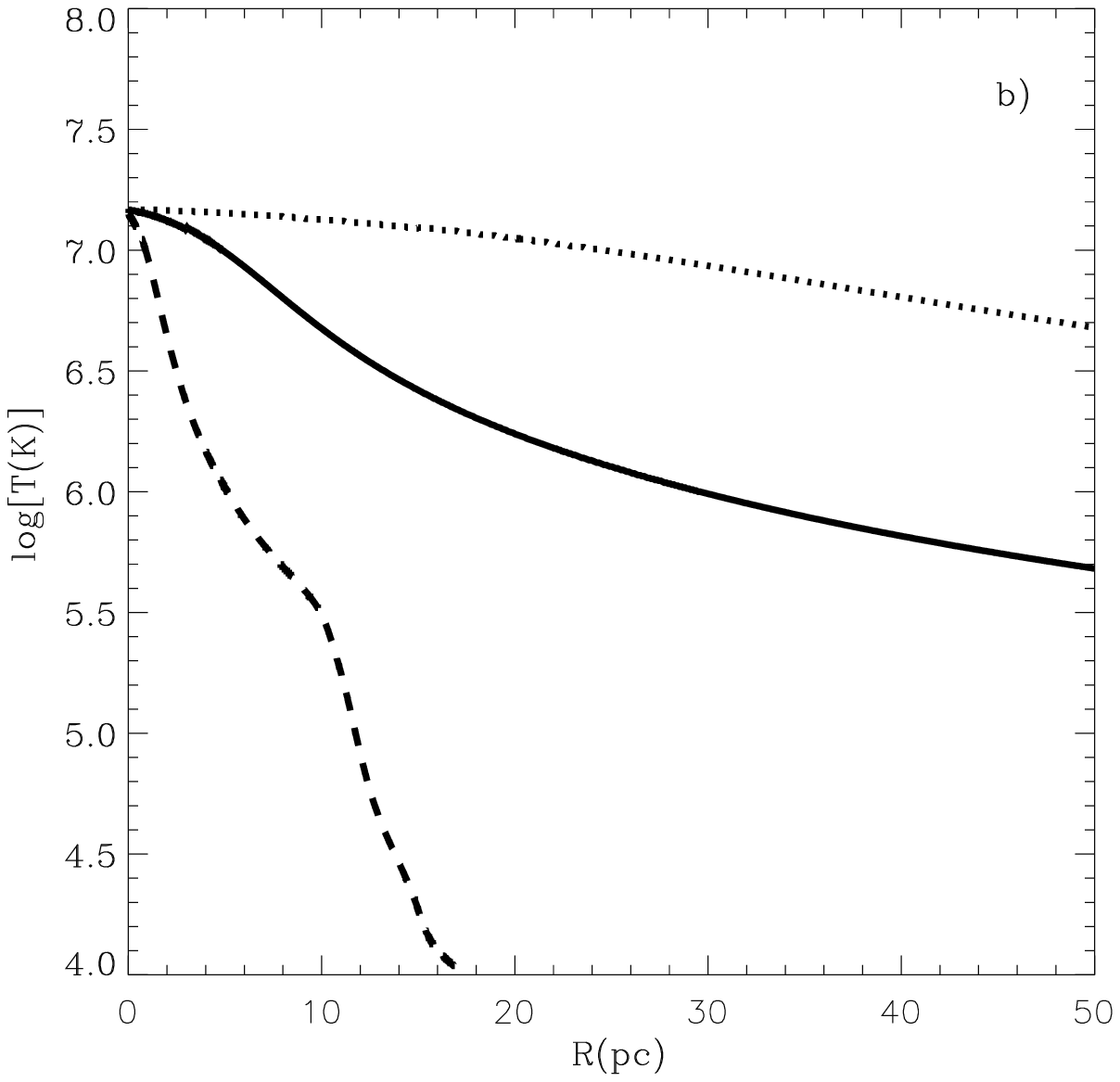}
\includegraphics{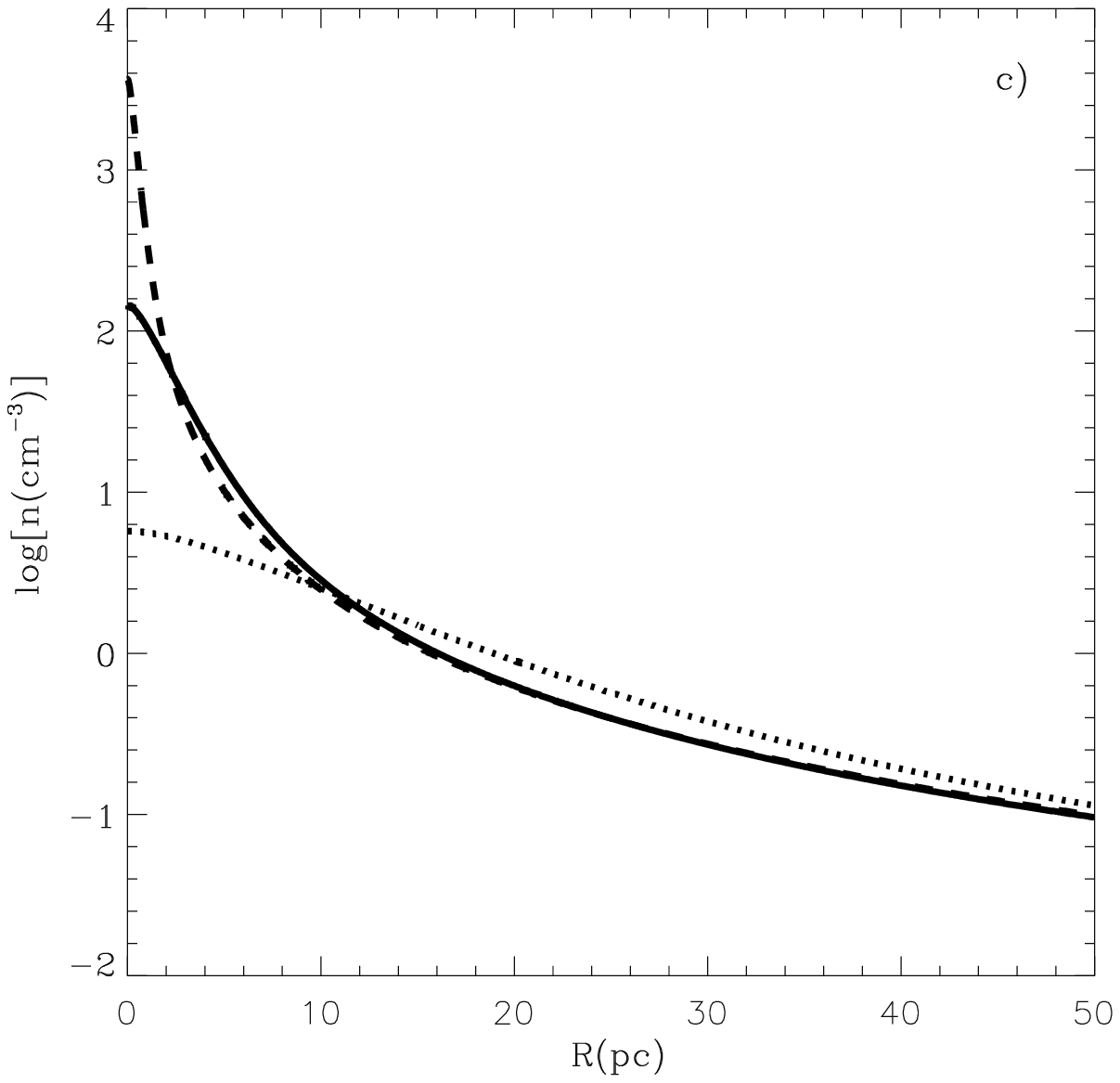}
\includegraphics{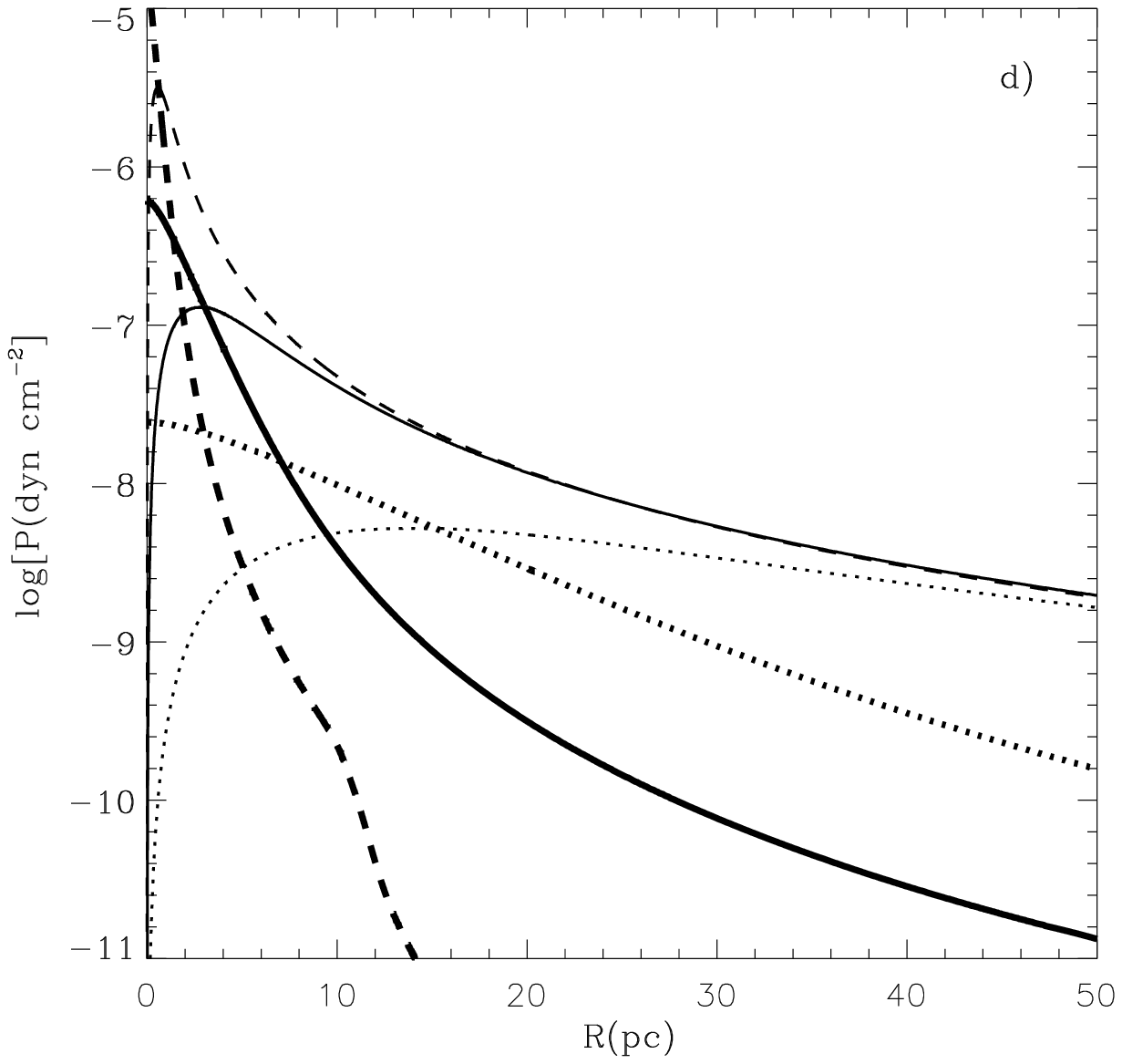}
\caption{The results of the calculations for models with
different $R_c$. Panels a, b, c and d present the
wind velocity, temperature, density and pressure, respectively. 
Solid, dotted and dashed lines present the results of the calculations
for models with $R_c = 1$~pc, $R_c = 5$~pc and $R_c = 0.2$~pc, respectively.
Thick and thin lines in panel d display the thermal and ram 
pressure, respectively.}
\label{fig4}
\end{figure}

\section{The catastrophic cooling regime}
\label{S7}

The density in the wind grows with the star cluster mechanical
luminosity/mass. It also increases if the wind is mass
loaded (as the adiabatic wind terminal speed is smaller) or if the cluster
is more compact (for a smaller $R_c$). In all these cases the impact of
radiative cooling on the flow hydrodynamics becomes progressively
more significant, as discussed in the previous section. The turn-off 
point in the temperature distribution moves towards the star cluster center
and the temperature rapidly drops to the $10^4$~K value.
At larger radii it falls even to lower values because of the gas expansion. 
Hereafter we will assume that the outflow is photoionized by young massive 
stars and thus the wind becomes isothermal as soon as the temperature reaches 
the $10^4$~K value. Outwards of this radius we replace the set of the main 
equations (\ref{eq7a})- (\ref{eq7c}) with equations describing isothermal
flows:
\begin{eqnarray}
 \label{eq11a}
      & & \hspace{-1.1cm} 
\der{u}{r}  = \frac{2 \rho c^2 / \gamma r - (1 + c^2/\gamma u^2) q_m u}
              {(1 - c^2/\gamma u^2) \rho u}
      \\[0.2cm] \label{eq11c}
      & & \hspace{-1.1cm}
\rho = \frac{2 q_{m0} R_c^3}{r^2 u} \left[1 - \left(1 +  \frac{r}{R_c} + 
       \frac{1}{2}\frac{r^2}{R^2_c}\right)
       \exp(-r/R_c)\right]  \, ,
\end{eqnarray}
where the sound speed, $c^2 = \gamma k T / \mu_a$, is constant and 
the temperature is $T = 10^4$~K. The thermal pressure then is:
$P = \rho c^2 / \gamma$. However, in the numeric code we obtain thermal 
pressure by integrating the differential equation
\begin{equation}
 \label{eq11b}
\der{P}{r} = \frac{c^2}{\gamma}\left(\frac{q_m}{u} - 
             \frac{\rho}{u}\der{u}{r} - \frac{2\rho}{r}\right) \, .
\end{equation}
This allows to minimize changes in the numerical code, when the
transition occurs from the radiative to the isothermal regime.

In the catastrophic cooling regime the singular point detaches from 
its quasi-adiabatic position and then rapidly moves towards the center. 
One can note that in this respect our solution is similar 
to that, found by \citet{1980MNRAS.191..711B}
for the accretion of the pre-heated gas onto a neutron star. In the later 
case the singular point is also located far away from the center, near the 
Bondi radius, if the accretion rate is low and the pre-heating of the 
accretion flow is small. However, when the accretion rate grows, the 
catastrophic heating regime sets in. The singular point detaches then from 
the Bondi radius and moves rapidly towards the neutron star surface.
  
Figure 5 presents an example of the catastrophic cooling regime. In 
this case the star cluster core radius and the adiabatic 
wind terminal speed are the same as in our reference model A, 
but the mechanical luminosity of the cluster is two orders of magnitude
larger: $L_{SC} = 3 \times 10^{42}$~erg s$^{-1}$.
\begin{figure}[htbp]
\vspace{18.0cm}
\includegraphics{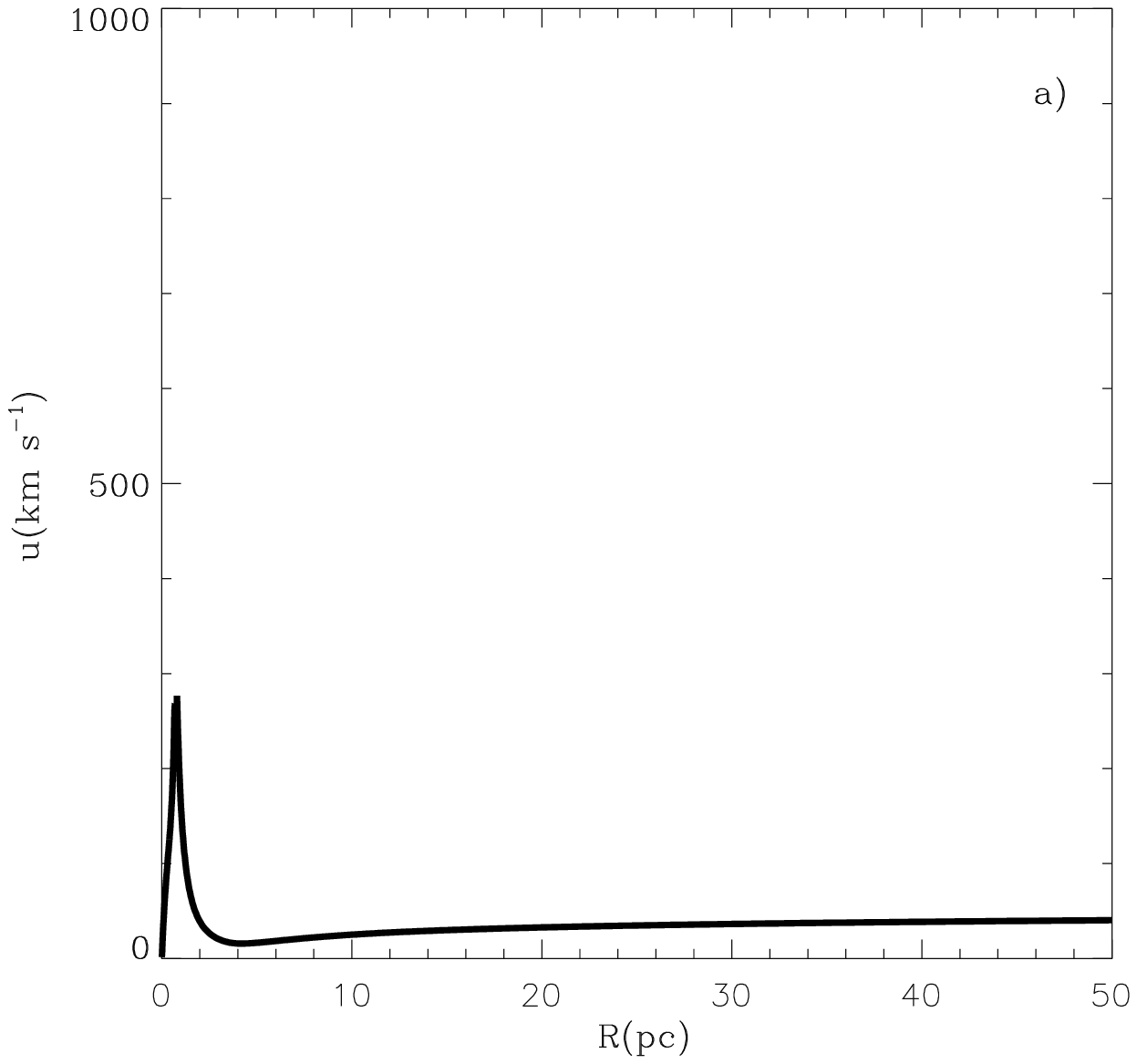}
\includegraphics{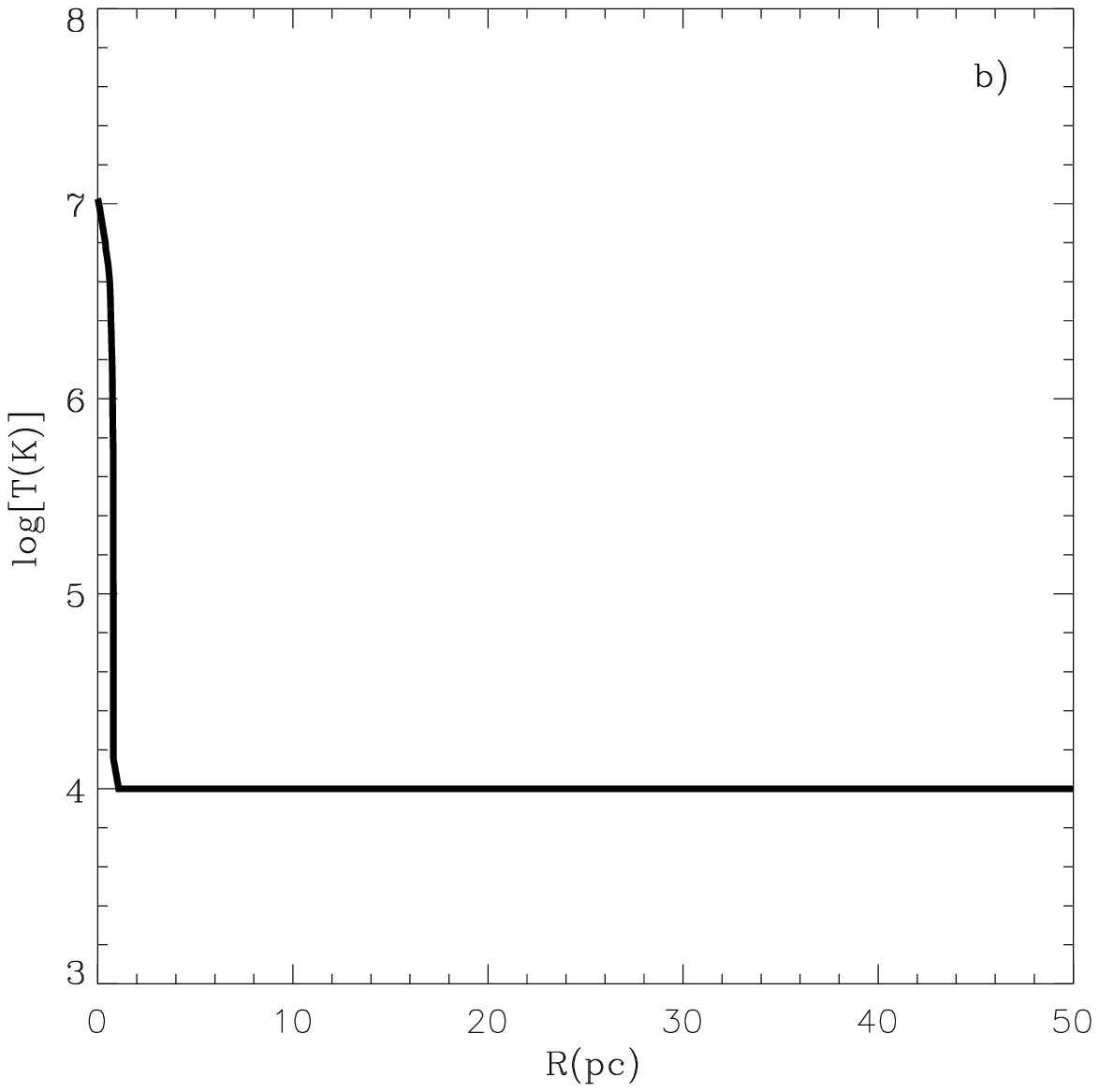}
\includegraphics{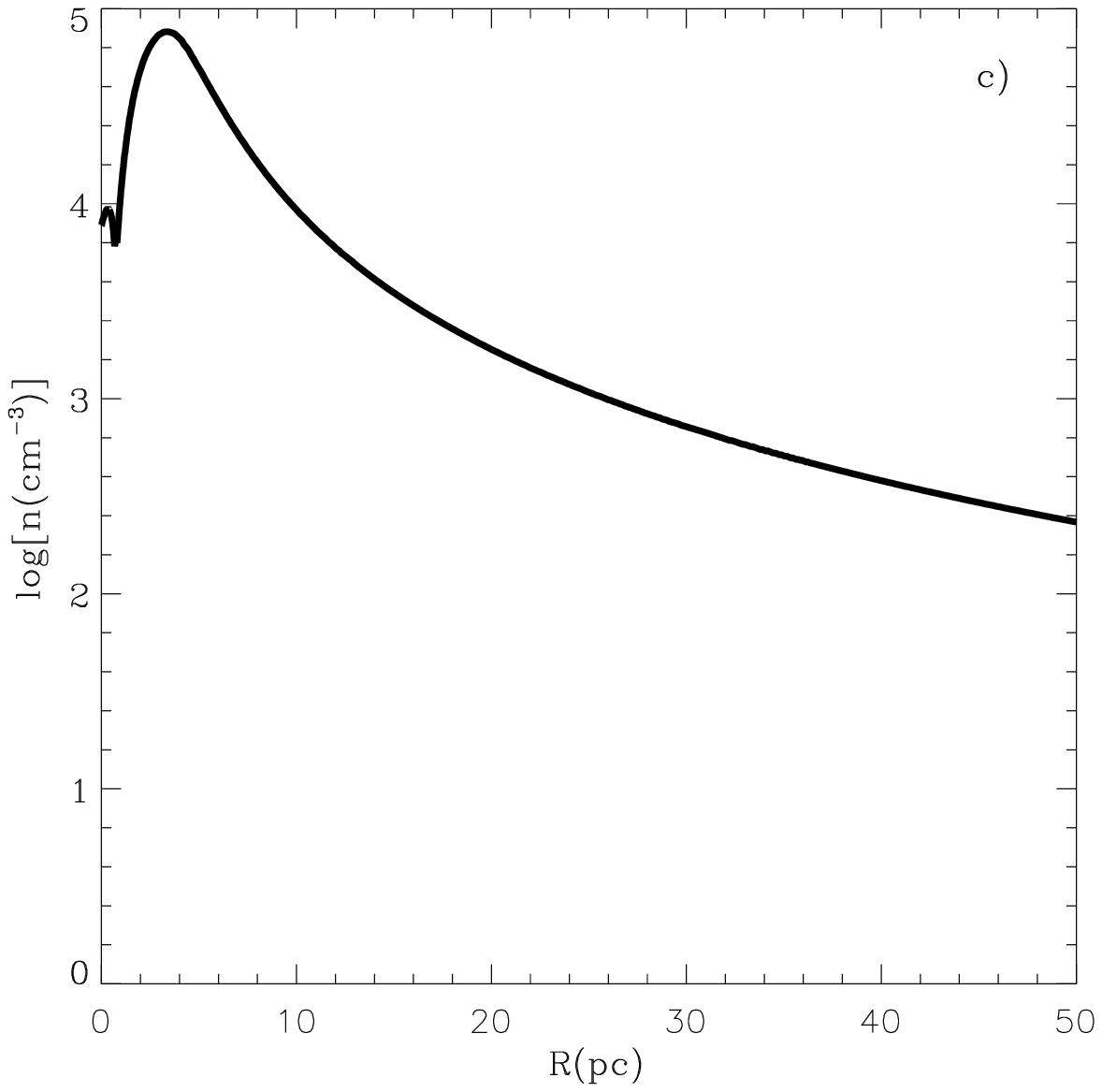}
\includegraphics{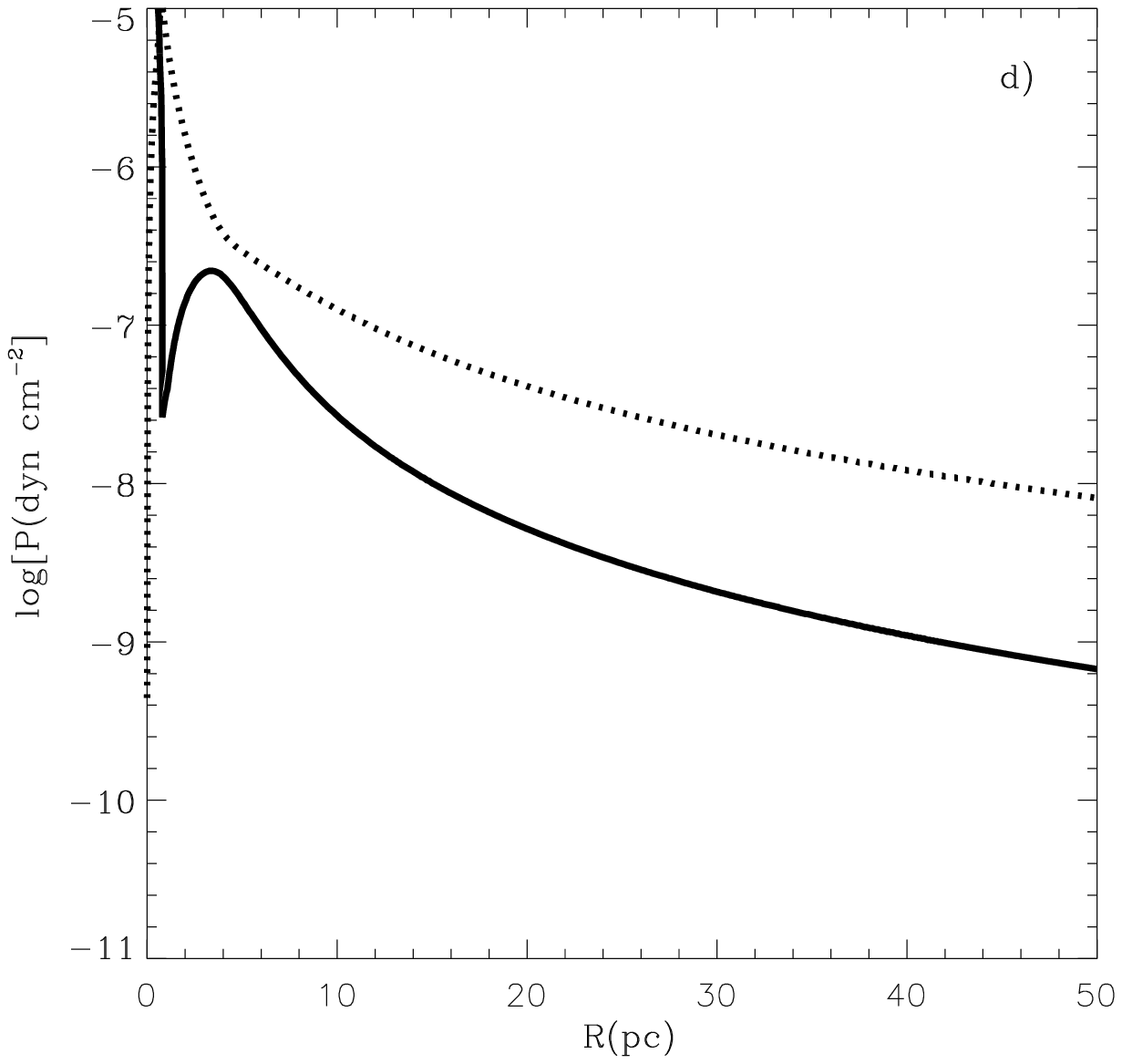}
\caption{Catastrophic cooling regime. The distribution of the 
hydrodynamic variables in the model with $L_{SC} = 3 \times 10^{42}$~erg 
s$^{-1}$, $R_c = 1$~pc and $V_{A\infty} = 1000$~km s$^{-1}$. Panels a, b, c 
and d present the wind velocity, temperature, density and pressure, 
respectively. Solid and dotted lines in panel d display the thermal and 
the ram pressure, respectively.}
\label{fig5}
\end{figure}
The singular point moves then inside the core radius, to 
$R_{sp} = 0.66$~pc distance from the star cluster center. The maximum flow 
velocity is much smaller, than in the quasi-adiabatic case.
It reaches only about 300~km s$^{-1}$ at the singular point (see panel a 
in Figure 5). The flow slows down then to about 16~km s$^{-1}$ being loaded
with the newly re-inserted matter with zero initial momentum.
In this regime the last term 
in equation (\ref{eq4b}) dominates over the thermal pressure gradient.
The velocity then increases slowly because in the isothermal regime the 
ionizing photons heat up and dump additional energy to the flow. The density 
grows in the region, where the wind slows down, and then decreases again 
in the isothermal wind region. The temperature drops abruptly to the $10^4$~K 
value when the flow passes the singular point. The thermal pressure also 
drops just after the singular point and then increases in the region where 
the density grows up and the temperature reaches a constant value. 
\begin{figure}[htbp]
\plotone{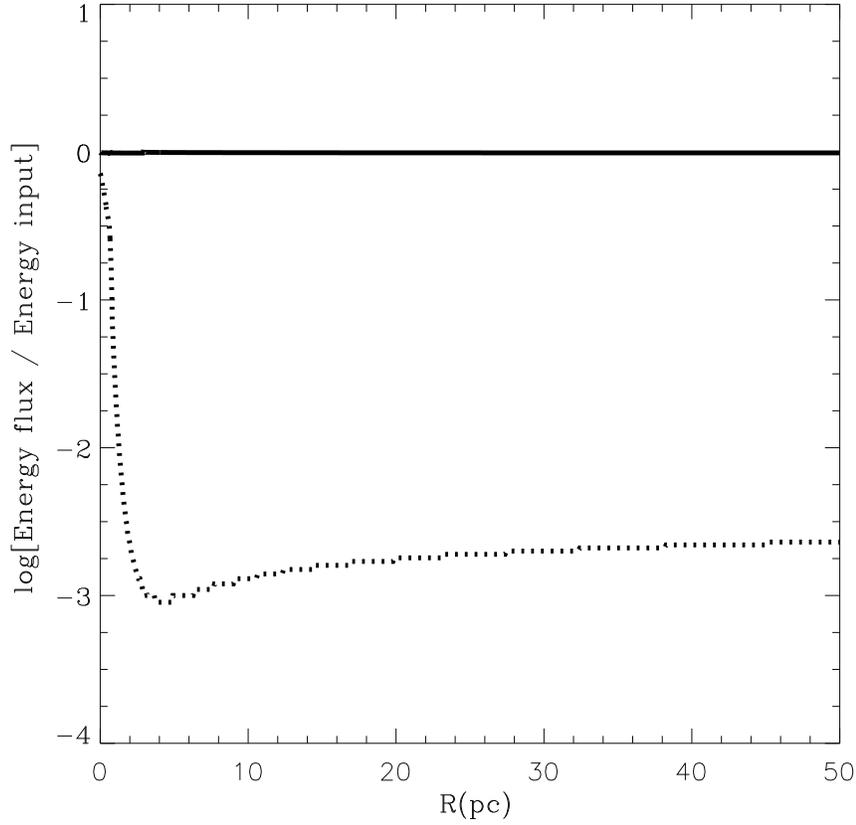}
\caption{The catastrophic cooling regime. The ratio of the mechanical energy 
         flux to the energy input rate as a function of radius. The solid 
         and dotted lines present the results of the calculations for
         the reference model A and the model with the same core radius 
         and adiabatic wind terminal speed, but two orders of magnitude 
         larger mechanical luminosity, respectively.}
\label{fig6}
\end{figure}

In the case under consideration the radiative losses of energy are 
catastrophic indeed.
Figure 6 shows the ratio of the energy flux through the surface with 
radius $r$: $L_w(r) = 4 \pi \rho u(r) r^2 (u^2(r)/2 + H(r))$, 
where $H$ is the enthalpy, to the mechanical energy input rate
inside the enclosed volume: $L_{SC}(r) = 
L_{SC} [1 - (1 + r/R_c + r^2/2R^2_c) \exp(-r/R_c)]$. 
The solid line in 
Figure 6 displays this ratio for the reference model A, whereas the
dotted one shows the same ratio in the catastrophic cooling regime.
In the catastrophic cooling regime $L_w(r)/L_{SC}(r)$ drops very rapidly
to about $10^{-3}$ value, whereas in the quasi-adiabatic model A
it is close to unity at any distance from the star cluster center.
This implies that in the extreme case, here discussed, the star cluster 
wind carries away only about 0.1\% of the deposited mechanical energy.  
The rest is radiated away because of strong radiative cooling. 

Note, that the catastrophic cooling regime also sets in if one
provides simulations for less energetic, but more compact clusters, 
and in the case of less energetic models with smaller adiabatic wind 
terminal speed parameter. In this regime the terminal wind 
velocity is small and may fall below the escape velocity. In this
case a fraction of the re-inserted matter might remain gravitationally 
bound and accumulate inside the star cluster volume. Thus, in the 
catastrophic cooling regime the gravitational pull from the cluster 
becomes an important factor \citep{2010ApJ...711...25S},
which should be 
included into the model. The impact that gravitational field of the 
cluster provides on the flow will be discussed in a future communication.    

\section{Comparison with homogeneous model predictions}
\label{S8}

In this section we confront the predictions from the exponential model with 
those, obtained under the assumption that stars are homogeneously distributed 
within the star cluster volume. Throughout this section we will assume that 
the two clusters have the same mass and their winds the same adiabatic 
terminal speed, but different, either exponential or homogeneous, stellar 
mass distributions. As shown in section \ref{S6} and in our prior papers, 
the distribution of the hydrodynamical variables and thus the observational 
manifistations of star cluster winds strongly depend on the characteristic 
space scale of the stellar mass distribution: the core radius, $R_c$, in 
models with an exponential stellar density distribution and on the star 
cluster radius, $R_{SC}$, in models with a homogeneous stellar distribution. 
Thus, one has to link these two parameters in order to compare the models. 
This could be done in different ways. For example, 
\citet{2006MNRAS.372..497J} compared 
two models assuming that in both cases the singular points are located at the 
same distance from the star cluster center. 
One can instead use the same half-mass radius $R_{hm}$ (e.g. 
\citealp{2010ARA&A..48..431P}).
Specifically, here we assume equally massive clusters with 
different stellar 
mass distributions but with the same half-mass radii: $R_{hme} = R_{hmh}$, 
where the half-mass radius is $R_{hme} = 2.67 R_c$ in the
exponential case and $R_{hmh} = 0.79 R_{SC}$ in models with a homogeneous 
mass distribution. The relation between the core radius $R_c$ and 
the star cluster radius $R_{SC}$ then is:
\begin{equation}
      \label{eq12}
R_c \approx 0.3 R_{SC} .
\end{equation}

\begin{figure}[htbp]
\vspace{17.0cm}
\includegraphics{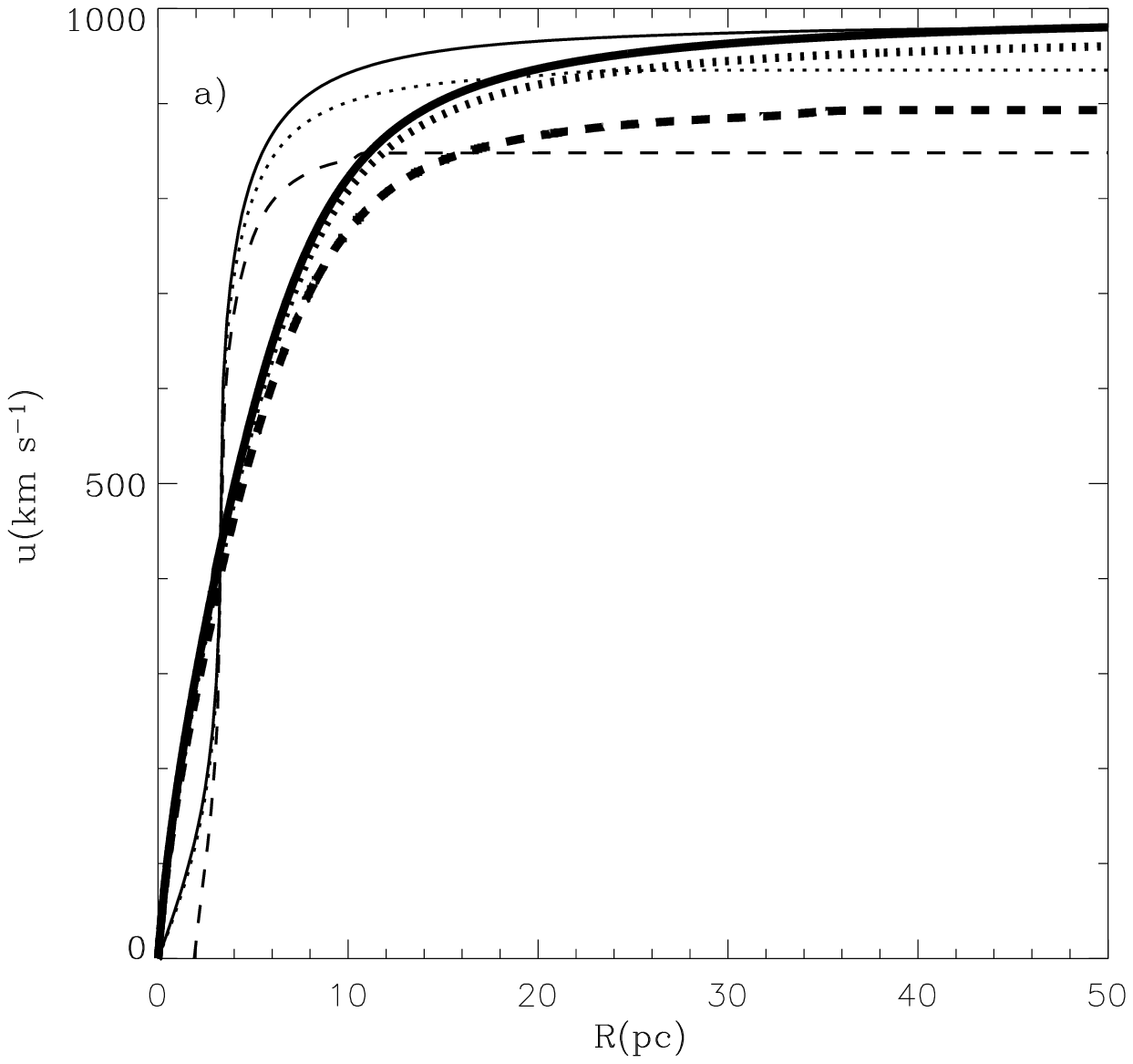}
\includegraphics{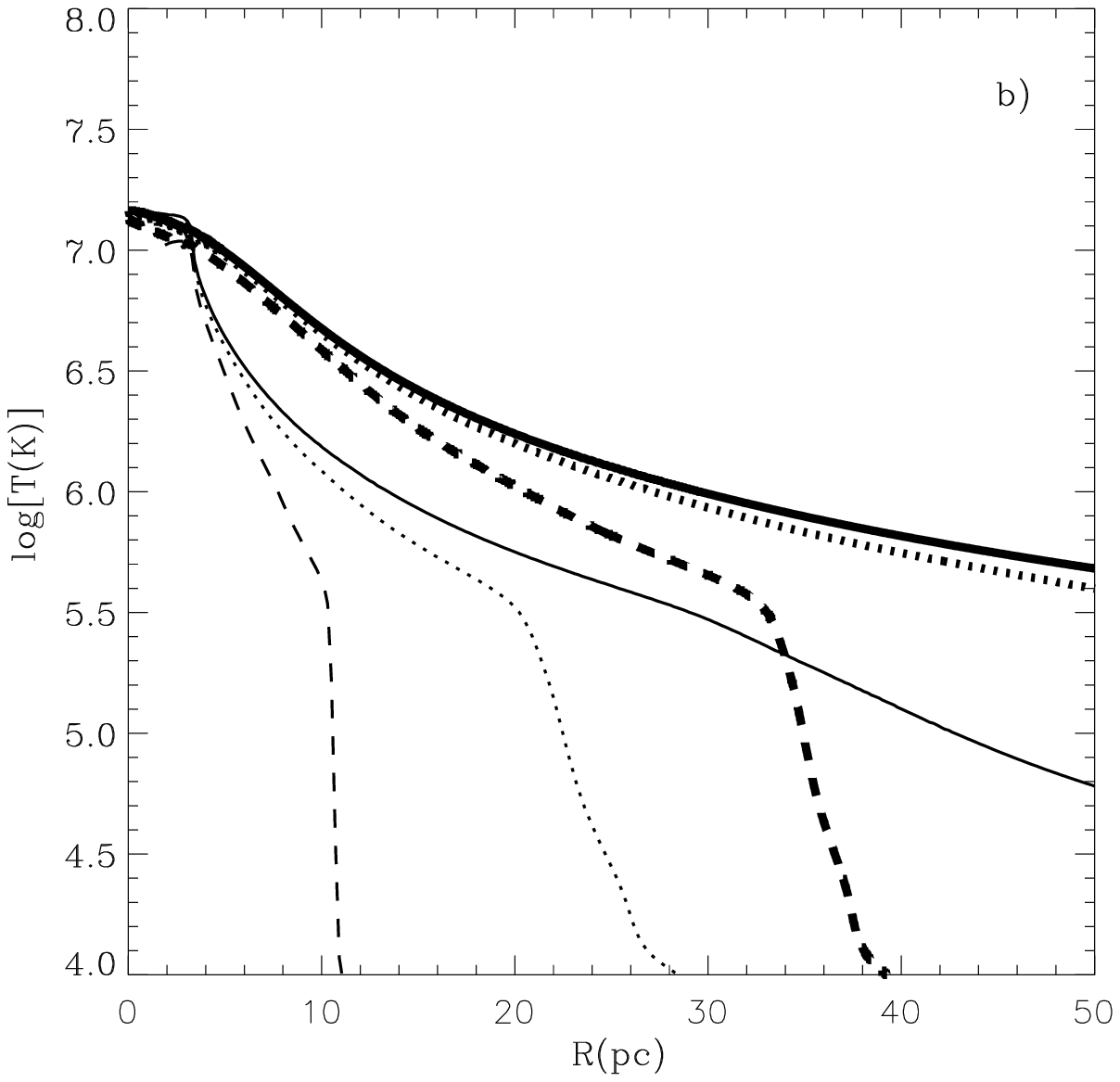}
\includegraphics{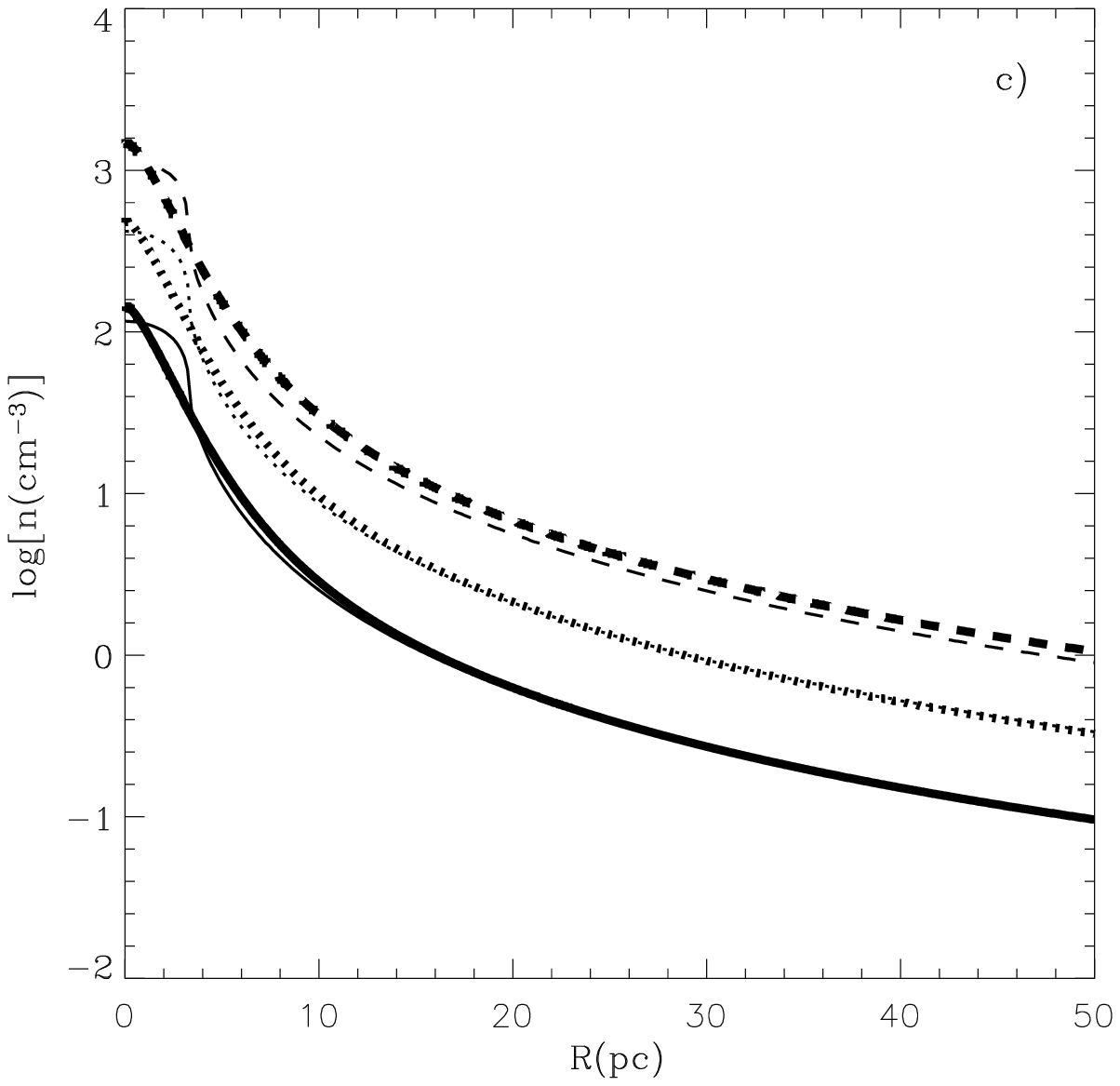}
\includegraphics{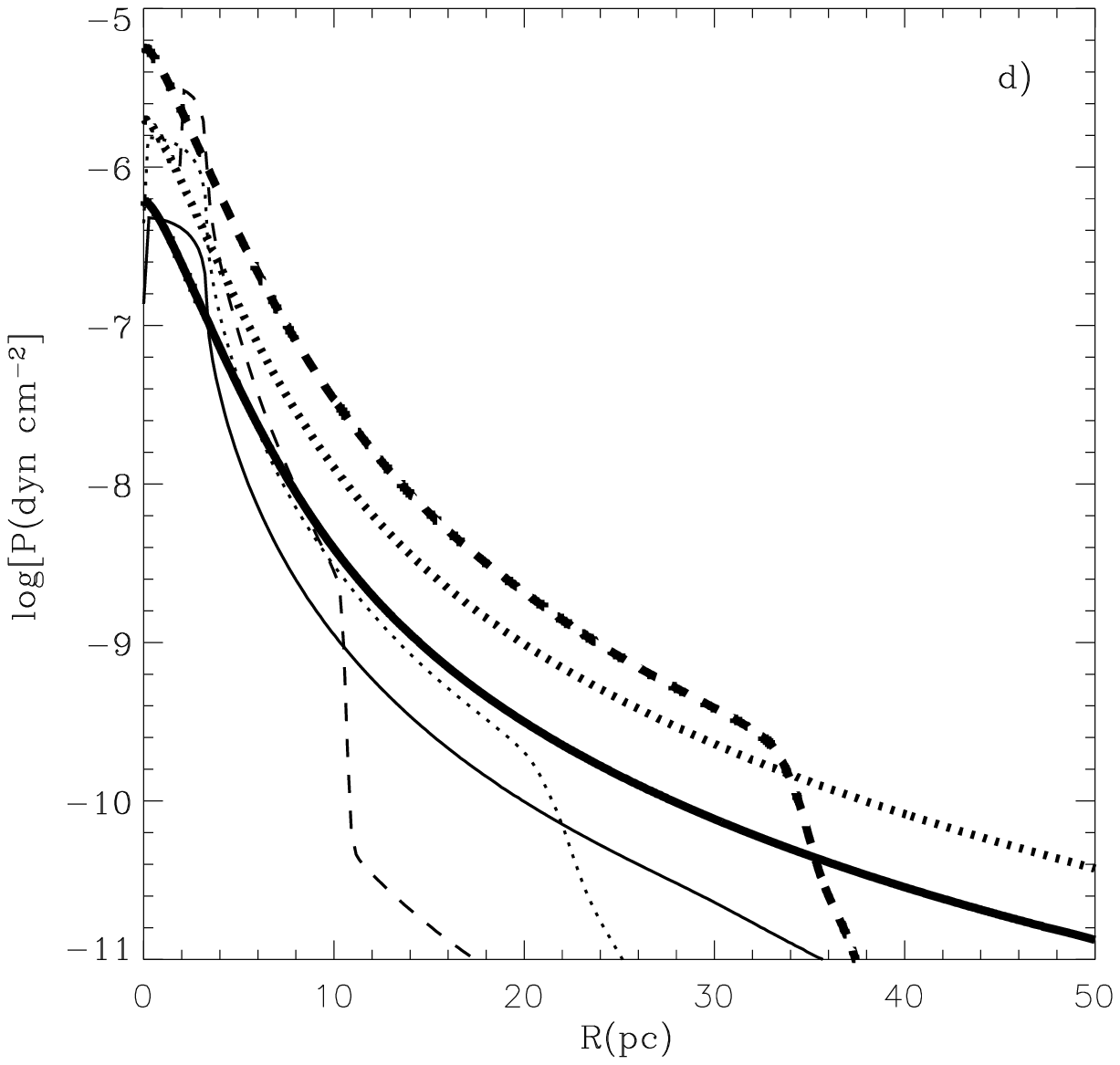}
\caption{The comparison of the exponential and the homogeneous model
predictions. Panels a, b, c and d present the distributions of the
wind velocity, temperature, density and thermal pressure, respectively.
Solid, dotted and dashed lines display the results of the calculations
for models with $L_{SC} = 3 \times 10^{40}$~erg s$^{-1}$, 
$L_{SC} = 10^{41}$~erg s$^{-1}$ and $L_{SC} = 3 \times 10^{41}$~erg s$^{-1}$.
Thick and thin lines show the distributions of the hydrodynamical variables in
the case with exponential and homogeneous mass distribution, respectively.
It was assumed that the adiabatic wind terminal speed parameter is the same
in all cases:  $V_{A\infty} = 1000$~km s$^{-1}$.}
\label{fig7}
\end{figure}
We present three cases: our reference models A and B and an intermediate 
model with an energy input rate $L_{SC} = 10^{41}$~erg s$^{-1}$ and 
assume that in each case the mass distribution may be either exponential, or 
homogeneous. The results of the calculations are presented in Figure 7.
Solid, dotted and dashed lines in Figure 7 display the results of the 
calculations for models with $L_{SC} = 3 \times 10^{40}$~erg s$^{-1}$, 
$L_{SC} = 10^{41}$~erg s$^{-1}$ and $L_{SC} = 3 \times 10^{41}$~erg s$^{-1}$,
respectively. Thick and thin lines show the distributions of the 
hydrodynamical variables in the case with an exponential and a homogeneous 
mass distribution, respectively. One can note, that models with an exponential 
stellar mass distribution are less affected by radiative losses of energy. 
Indeed, in the calculations with homogeneous mass distribution the temperature
and the thermal pressure already deviate significantly from the 
quasi-adiabatic profiles when the star cluster mechanical luminosity is 
$L_{SC} = 10^{41}$~erg s$^{-1}$ whereas in the exponential case are not 
(compare thin and thick dotted lines in panels b and d). The mechanical 
energy input rate in the most energetic  homogeneous model with 
$L_{SC} = 3 \times 10^{41}$~erg s$^{-1}$ exceeds the threshold value 
(see Figure 2 in \citealp{2007ApJ...658.1196T}).
In this case the stagnation 
point (the point where the wind velocity is 0 km s$^{-1}$) shifts from the 
center to $R_{st} = 1.9$~pc and the shock-heated plasma becomes thermally 
unstable within the central zone with $r \le R_{st}$ 
\citep{2008ApJ...683..683W}.
We did not find a similar bimodal regime in calculations with an 
exponential stellar distribution. In these cases, the faster drop 
in density inhibits catastrophic cooling in the center. 
The two models are quite different in this respect.  In the models with
a homogeneous star distribution, the position of the singular point is fixed 
at $r = R_{SC}$ but the stagnation point may move from the center due to 
catastrophic cooling. In the models with an exponential stellar mass 
distribution it is quite the opposite: the stagnation point remains always
at the center, whereas the singular point may detach from its quasi-adiabatic 
position and move closer towards the center. Thus the definition of the 
threshold mechanical luminosity as the mechanical luminosity above which 
the stagnation point moves from the star cluster center and the central 
zone becomes thermally unstable does not occur in models with an exponential 
stellar mass distribution. However, the flow may be thermally unstable 
outside of the singular point in this case. This will be thoroughly discussed 
in a forthcoming communication.
\begin{figure}[htbp]
\plotone{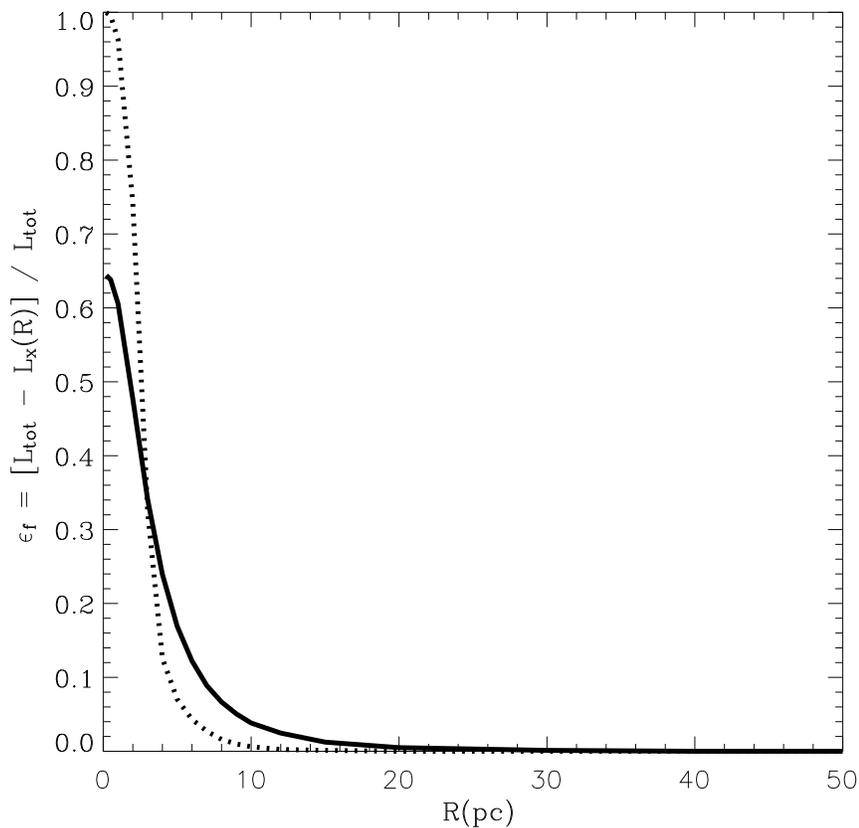}
\caption{The distribution of the X-ray emission along the star cluster
         wind. The solid and dotted lines show the distribution of the 
         X-ray luminosity $\epsilon_X$ (see the text) along the star 
         cluster wind in the case of the exponential and the homogeneous
         stellar mass distribution, respectively. Note, that both
         luminosities are normalized to the total homogeneous wind 
         luminosity, $L_{Xtot} = 8.1 \times 10^{38}$~erg s$^{-1}$, and
         that X-ray emission is slightly more concentrated in the case with 
         homogeneous stellar mass distribution.}
\label{fig8}
\end{figure}

Diffuse X-ray emission has been detected from many young stellar clusters
and their associated HII regions (e.g. \citealp{2002ApJ...573..191M,
2004ApJ...611..858L}, 
see the review of the resent results in \citealp{2011ApJS..194...16T}).
\citet{2000ApJ...536..896C}, \citet{2001ApJ...559L..33R}, 
\citet{2003MNRAS.339..280S}, \citet{2005ApJ...635.1116S},
\citet{2005ApJ...623..171R} and \citet{2007MNRAS.380.1198R}
suggested that the observed diffuse X-ray emission manifests the hot, 
shock-heated star cluster winds. The contribution from the hot massive stars
to the observed X-ray emission has been discussed by 
\citet{2005MNRAS.361..679O}.
The X-ray luminosity of the star cluster wind then is:
\begin{equation}
      \label{eq13}
L_X = 4 \pi \int_0^{R_{out}} r^2 n_e n_i \Lambda_X(T,Z) {\rm d}r ,
\end{equation}
where $n_e(r)$ and $n_i(r)$ are the electron and ion number densities, 
$\Lambda_X(Z,T)$ is the X-ray emissivity used by 
\citet{2000MNRAS.314..511S} and $R_{out}$ marks either the location of 
the outer wind driven shock,
or the X-ray cut-off radius (the radius where the temperature in the
wind drops below $T_{cut} \approx 5 \times 10^5$K).  We integrate equation
(\ref{eq13}) numerically using the temperature and density profiles 
obtained from calculations with either exponential or homogeneous stellar
mass distribution, assuming that $n_e = n_i = \rho(r) / \mu_i$,
where $\mu_i = 14/11 m_H$ is the ion number density. We found that the
exponential model predicts a slightly smaller (within a factor of two)
X-ray luminosity. For example, when the mechanical luminosity of the
cluster is  $L_{SC} = 3 \times 10^{40}$~erg s$^{-1}$ (model A), the
calculations with exponential stellar mass distribution predict the total
0.3 keV - 8.0 keV wind luminosity $L_{Xtot} = 5.2 \times 10^{38}$~erg s$^{-1}$
whereas the homogeneous model leads to $L_{Xtot} = 8.1 \times 
10^{38}$~erg s$^{-1}$. When  $L_{SC} = 10^{41}$~erg s$^{-1}$, the exponential 
model predicts $L_{Xtot} = 6.1 \times 10^{39}$~erg s$^{-1}$ whereas the 
homogeneous one $L_{Xtot} = 1.0 \times 10^{40}$~erg s$^{-1}$.  We cannot 
compare the X-ray luminosities in the most energetic case C because the 
central zone in the model with homogeneous stellar mass distribution is 
thermally unstable and the distributions of the hydrodynamical variables 
inside this zone cannot be obtained in the  semi-analytic calculations. 
Figure 8 compares the distributions of the X-ray emission along the wind, 
$\epsilon_X = [L_{Xtot} - L_X(r)] / L_{Xtot}$, in the case when the star
cluster mechanical luminosity is $L_{SC} = 3 \times 10^{40}$~erg s$^{-1}$. 
Very similar results were obtained in the calculations where instead of 
$R_{hme} = R_{hmh}$ the same singular radius, as suggested by 
\citet{2006MNRAS.372..497J} was used for the two stellar mass distribution 
models.

\section{Summary}

Here we present, for the first time, a radiative semi-analytic solution for 
steady state, spherically-symmetric winds driven by stellar clusters with 
an exponential stellar density distribution. The method, here developed,
improves previous calculations provided for stellar clusters with a given 
size and a homogeneous stellar density distribution and thus leads to more 
reliable hydrodynamic predictions. It may be easily extended to clusters with 
other stellar density distributions.

In our model, unlike in most previous calculations, the position of the 
singular point, $R_{sp}$, where the transition from the subsonic to the 
supersonic flow occurs, is not associated with the star cluster edge, 
but calculated from the condition that the integral curve must pass 
through the singular point. When radiative losses of energy are 
negligible, the singular radius is always about $R_{sp} \approx 4 R_c$, where 
$R_c$ is the star cluster core radius, irrespective of the other star cluster 
parameters. This is not the case in the catastrophic cooling regime, when the 
temperature drops abruptly at a short distance from the star cluster center 
and the transition from the subsonic to the supersonic regime occurs at the 
much smaller distance from the star cluster center.

Radiative cooling becomes a significant factor when the cluster is very 
energetic/massive, compact, or the adiabatic wind terminal speed parameter,
$V_{A\infty} = (2 L_{SC} / {\dot M}_{SC})^{1/2}$, is small. 
In the catastrophic cooling regime outflows carry away of the star cluster 
region only a small fraction of the deposited mechanical energy. The 
gravitationally bound, partially ionized nebulae may be formed then, if the 
photoionized gas cannot escape the gravitational well of the cluster. 
On the other hand, the low mass clusters with small energy input rates and 
large radii drive quasi-adiabatic winds. In these cases our results show 
an excellent agreement with the results of non-radiative 1D numerical 
simulations. 

The star cluster driven wind model presented here may be applied to 
many problems, which are currently discussed in the literature. Such,
as the star cluster diffuse X-ray emission, the origin of compact HII 
regions, which are frequently detected around young massive clusters 
and the origin of the low-ionization line emission in the starburst 
driven galactic-scale outflows. We will address some of them in a 
future communication.
 
\acknowledgments

Our thanks to an anonymous referee for a rapid report and many important 
suggestions that have greatly improved the paper and to Prof. J. Palou\v{s}
and Dr. R. W\"unsch for their comments regarding the flow thermal stability. 
We also appreciate the friendly atmosphere of the 13th edition of
the Guillermo Haro Workshop in Tonantzintla (M\'exico), where
the wind model for stellar clusters with an exponential stellar density
distribution was discussed for the first time. This study 
has been supported by CONACYT - M\'exico, research grant 131913.
GBK has been partially supported by Russian Foundation for Basic 
Research, grants 08-02-00491 and 11-02-00602, the RAN Program 
``Origin, formation and evolution of objects in the Universe'' and
Russian Federation President Grant for Support of Leading Scientific
Schools NSh-3458.2010.2.

\appendix
\renewcommand{\theequation}{A\arabic{equation}}

\section*{Appendix}

In order to obtain the flow velocity, pressure, their derivatives, and also 
the density and the temperature at the singular point, one has to use the 
condition that at this point the numerator and the denominator in 
equation (\ref{eq7a}) vanish. The denominator in equation (\ref{eq7a})
vanishes when  the wind velocity reaches the local speed of sound and thus:  
$u_{sp} = c_{sp}$.
The density in the singular point (see equation \ref{eq7c}) then is:
\begin{equation}
      \label{A1}
\rho_{sp} = \frac{2 q_{m0} R_c^3}{R^2_{sp} c_{sp}} 
            \left[1 - \left(1 +  \frac{R_{sp}}{R_c} + 
            \frac{1}{2}\frac{R_{sp}^2}{R^2_c}\right)
            \exp(-R_{sp}/R_c)\right]  \, 
\end{equation}
The second condition, that the numerator in equation (\ref{eq7a})  vanishes, 
then yields:
\begin{equation}
      \label{A2}
c^4_{sp} - 2 F_1(R_{sp}) c^2_{sp} + F_2(R_{sp}) \Lambda(T_{sp}, Z) = 0 \, ,
\end{equation}
where functions $F_1$ and $F_2$ are: 
\begin{eqnarray}
 \label{A3a}
      & & \hspace{-1.1cm} 
F_1  = \frac{(\gamma-1)}{4 F_3(R_{sp})} V^2_{A\infty} \exp(-R_{sp}/R_c) \, ,
      \\[0.2cm]     \label{A3b}
      & & \hspace{-1.1cm}
F_2 = \frac{4 (\gamma-1) q_{m0} R_c^6}{\mu^2 _i R^4_{sp} F_3(R_{sp})} 
      \left[1 - \left(1 +  \frac{R_{sp}}{R_c} + 
      \frac{1}{2}\frac{R_{sp}^2}{R^2_c}\right)
      \exp(-R_{sp}/R_c)\right]^2 \, ,
\end{eqnarray}

and 

\begin{eqnarray}
\label{A3c}
     & & \hspace{-1.9cm}
F_3 = 4 \left(\frac{R_c}{R_{sp}}\right)^3 (1 - \exp(-R_{sp}/R_c)) - 
                                                                  \nonumber
      \\[0.2cm]     
      & & \hspace{-0.0cm}
         \left[\frac{\gamma+1}{2} + 
      4 \left(\frac{R_c}{R_{sp}}\right)^2 \left(1 + 
      \frac{R_{sp}}{2 R_c}\right)\right] \exp(- R_{sp}/R_c) \, ,
\end{eqnarray}
$\mu_i = 14/11 m_H$ is the mean mass per ion.

This nonlinear algebraic equation defines the temperature at the singular
point, $T_{sp}$, if $R_{sp}$ is known. One can present equation
(\ref{A2}) in the dimensionless form and then solve it numerically:
\begin{equation}
      \label{A4}
FA = 1 - 2 F_1(R_{sp}) c^{-2}_{sp} + 
     F_2(R_{sp}) \Lambda(T_{sp}, Z) c^{-4}_{sp} = 0 .
\end{equation}
Equation (\ref{A4}) may have one, two, or have nor real roots as it is 
shown in Figure 9, which displays function FA tabulated at $R_{sp} = 4$~pc 
radius for three different values of the star cluster mechanical luminosity: 
$L_{SC} = 3 \times 10^{40}$~erg s$^{-1}$, $L_{SC} = 
3 \times 10^{41}$~erg s$^{-1}$ and $L_{SC} = 3 \times 10^{42}$~erg s$^{-1}$ -
solid, dotted and dashed lines, respectively.
The proper solution of equation (\ref{A4}) is selected from the condition, that
segments of the integral curve obtained by the outward integration from the 
star cluster center and by the inward integration from the singular point 
match in an interior radius $0 < R_{fit} < R_{sp}$. Note, that one has to 
obtain the position of the singular point, $R_{sp}$, by iterations as it is 
described in Section 5.

Having the value of $T_{sp}$, one can obtain the velocity in the singular 
point, which is: $u_{sp} = c_{sp}$. The 
density in the singular point yields from equation (\ref{A1}), the pressure 
then is: $P_{sp} = \rho_{sp} c^2_{sp} / \gamma$. 
Thus, one can obtain the values of all hydrodynamic variables 
in the singular point solving the nonlinear algebraic equation (\ref{A4}). The 
value of the singular radius, $R_{sp}$, is obtained by iterations, as
explained in section \ref{S5}. 

\begin{figure}[htbp]
\plotone{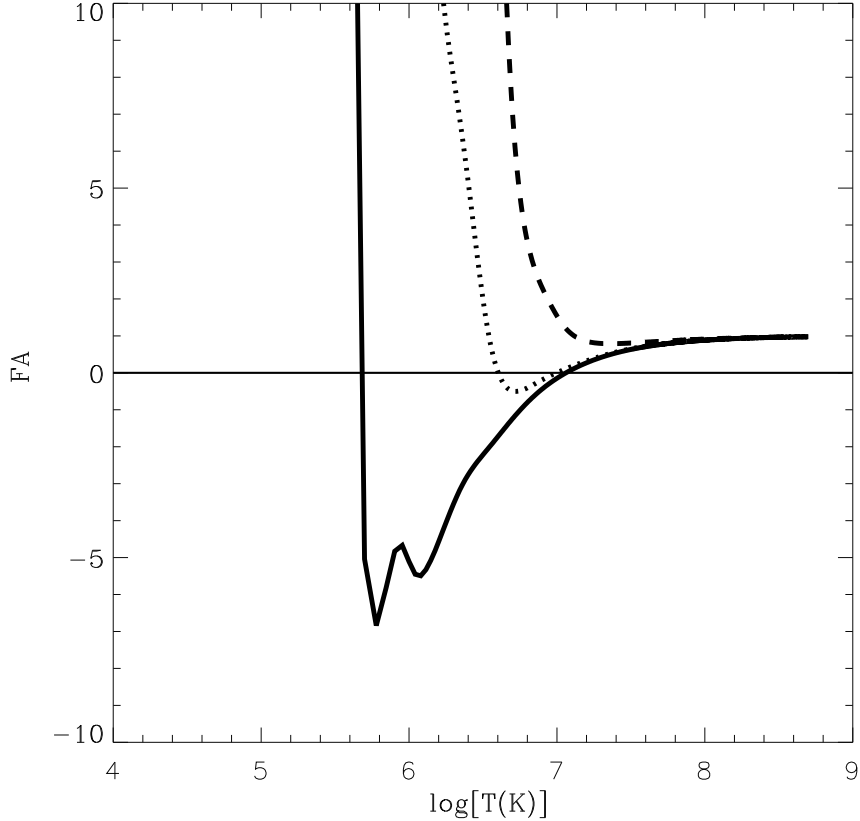}
\caption{Different roots of equation (\ref{A4}). Solid, dashed and dotted 
lines display function FA tabulated at $R_{sp} = 4$~pc for three different  
mechanical luminosities: $L_{SC} = 3 \times 10^{40}$~erg s$^{-1}$, $L_{SC} = 
3 \times 10^{41}$~erg s$^{-1}$ and $L_{SC} = 3 \times 10^{42}$~erg s$^{-1}$,
respectively.}
\label{fig9}
\end{figure}

In order to obtain the derivative of velocity in the singular point, one can 
use the L'Hopital's rule. The derivatives of numerator and denominator of
equation (\ref{eq7a}) over radius are:
\begin{eqnarray}
 \label{A5a}
      & & \hspace{-2.1cm} 
\der{N}{r}  = \derp{N}{r} + \derp{N}{u} \der{u}{r} + \derp{N}{c^2} \der{c^2}{r} +
                      \derp{N}{T} \der{T}{r} + \derp{N}{\rho} \der{\rho}{r} =
                      F_5 \der{u}{r} + F_6 + \derp{N}{r} \, ,                      
      \\[0.2cm]     \label{A5b}
      & & \hspace{-2.1cm}
\der{D}{r} = \derp{D}{u} \der{u}{r} + \derp{D}{c^2} \der{c^2}{r} =
                     -(\gamma + 1) c \left(\der{u}{r} + 
             \frac{q_m}{\rho}\right) + \frac{2 c^2}{r} \, ,
\end{eqnarray}
where functions $N$, $D$, $\difp{N}/\difp{r}$, $F_5$ and $F_6$ are:
\begin{eqnarray}
 \label{A6a}
      & & \hspace{-1.1cm} 
N(r, u, \rho, c, T)  = (\gamma-1)(q_e - Q) - 4 q_{m0} c^2 (R_c/r)^3 +
                                                                  \nonumber
      \\[0.2cm]     
      & & \hspace{-0.0cm}
       q_m [(\gamma+1) u^2 / 2 + 4 c^2 (R_c/r)^3 (1 + r/R_c + (r/R_c)^2 / 2)] 
       \, ,
      \\[0.2cm]     \label{A6b}
      & & \hspace{-1.1cm}
D(u, c) = c^2 - u^2  \,  
      \\[0.2cm]     \label{A6f}
      & & \hspace{-1.1cm}
\derp{N}{r} = - \frac{1}{R_c} \left[(\gamma-1) q_e + 
      4 q_{m0} c^2 \left(\frac{R_c}{r}\right)^3
      \left[\exp(-r/R_c) - \frac{3 R_c}{r}(1 - \exp(-r/R_c))\right]\right. +
                                                                \nonumber
      \\[0.2cm]     
      & & \hspace{-0.0cm}
  \left. 4 q_m c^2 \left[\frac{\gamma+1}{8} + 
   2 \left(\frac{R_c}{r}\right)^3 + \frac{3}{2} \left(\frac{R_c}{r}\right)^2 + 
  \frac{1}{2}\frac{R_c}{r} \right]\right] \, ,
  \\[0.2cm]  \label{A6c}
      & & \hspace{-1.1cm} 
F_5 = (1 - \gamma) c F_4 + (1 + \gamma) q_m c + 2 (\gamma - 1) \rho^2 \Lambda /
          c \mu^2_i
      \\[0.2cm]     \label{A6d}
      & & \hspace{-1.1cm}
F_6  = \frac{2 (\gamma - 1) \Lambda \rho^2}{\mu^2_i c} \left(\frac{2c}{r} - 
            \frac{q_m}{\rho}\right) - \left[\frac{(\gamma+1) q_m c}{\rho} -
            \frac{2 c^2}{r}\right] F_4 \, ,
\end{eqnarray}

and

\begin{eqnarray}
\label{A6e}
      & & \hspace{-1.9cm}
F_4  = \frac{1-\gamma}{\gamma} \frac{\rho^2}{\mu_i k} \derp{\Lambda}{T} +
                                                                \nonumber
      \\[0.2cm]     
      & & \hspace{0.0cm}
       4 q_{m0} \left(\frac{R_c}{r}\right)^2 \left[\left(1 + 
       \frac{r}{2 R_c}\right)\exp(-r/R_c) - 
       (1 - \exp(-r/R_c)) \frac{R_c}{r}\right] \, .
\end{eqnarray}

One can obtain then the derivative of the wind velocity (and thus the 
derivative of the thermal pressure) at the singular point substituting 
relations (\ref{A5a}) and (\ref{A5b}) into equation (\ref{eq7a}) and 
keeping in mind that at the singular point $u_{sp} = c_{sp}$. This leads to 
a quadratic algebraic equation:
\begin{equation}
      \label{A8}
\left(\der{u}{r}\right)^2 - 2 F_7 \der{u}{r} + F_8 = 0 \, ,
\end{equation}
where functions $F_7$ and $F_8$ are:
\begin{eqnarray}
 \label{A9a}
      & & \hspace{-1.1cm} 
F_7  = \left[2 \rho c^2 / r - (\gamma+1) c q_m - F_5\right] / 2 (\gamma+1) c \rho \, ,
      \\[0.2cm]     \label{A9b}
      & & \hspace{-1.1cm}
F_8 = [F_6 + \difp{N}/\difp{r}] / [(\gamma+1) c \rho] \, .
\end{eqnarray}
The root of equation (\ref{A8}), which leads to the positive derivative 
$\dif{u}/\dif{r}$ at the singular point, is used in the calculations.

\bibliographystyle{aa}
\bibliography{ewind}

\end{document}